\documentclass{aa}  

\usepackage{graphicx}
\usepackage{txfonts}
\usepackage{lipsum}
\usepackage{subcaption}         
\usepackage{lscape}             
\usepackage{placeins}           

\usepackage{sidecap}

\def\tari{IRAS~21078$+$5211}
\def\targ{G035.02$+$0.35}

\def\nh3{NH$_{3}$}
\def\kms{km~s$^{-1}$}

\def\Vlsr{$V_{\rm LSR}$}
\def\Jyb{Jy~beam$^{-1}$}

\def\G24{G24.78$+$0.08}

\newcommand{\ms}{$M_{\odot}$}

\newcommand{\pas}{$\rlap{.}^{\prime\prime}$}
\newcommand{\pss}{$\rlap{.}^{\rm s}$}

\newcommand{\degree}{$^{\circ}$}

\begin{document}

   \title{Protostellar Outflows at the EarliesT Stages (POETS)} \subtitle{IX. Magnetohydrodynamic disk winds traced by SO and SO$_2$ in luminous protostars}

   \titlerunning{SO and SO$_2$ tracing MHD DWs}

   \author{L. Moscadelli\inst{1}
        \and 
        H. Beuther\inst{2}
        \and
        A. Sanna\inst{3}
        \and 
        M.~T. Beltr{\'a}n\inst{1}
        \and
        C. Gieser\inst{2}
         \and
         Th. Henning\inst{2}
         \and
         P.~D. Klaassen\inst{4}
         \and 
         R. Kuiper\inst{5}
         \and
         S. Leurini\inst{3}
         \and
         T. M\"oller\inst{6}
         \and
         A. Palau\inst{7} 
         \and 
         R.~E. Pudritz\inst{8}
          \and
          \'A S{\'a}nchez-Monge\inst{9,10}
         \and
         D. Semenov \inst{11,2}
          \and
         J.~S. Urquhart\inst{12}
         \and
         H. Zinnecker \inst{13}}

   \institute{INAF - Osservatorio Astrofisico di Arcetri, Largo E. Fermi 5, I-50125, Firenze, Italy
             \email{luca.moscadelli@inaf.it}
             \and
          Max-Planck-Institut für Astronomie, Königstuhl 17, 69117 Heidelberg, Germany
             \and 
       INAF - Osservatorio Astronomico di Cagliari, Via della Scienza 5, 09047 Selargius (CA), Italy
             \and 
  UK Astronomy Technology Centre, Royal Observatory Edinburgh, Blackford Hill, Edinburgh, EH9 3HJ, UK
            \and 
           Faculty of Physics, University of Duisburg-Essen, Lotharstra{\ss}e 1, D-47057 Duisburg, Germany
           \and 
          I.\ Physikalisches Institut, Universit\"at zu K\"oln, Z\"ulpicher Str.\ 77, D-50937 K\"oln, Germany 
           \and 
Universidad Nacional Aut\'onoma de M\'exico, Instituto de Radioastronom\'ia y Astrof\'isica, Antigua Carretera a P\'atzcuaro 8701, Ex-Hda. San Jos\'e de la Huerta, 58089, Morelia, Michoac\'an, M\'exico
           \and
           Department of Physics and Astronomy, McMaster University, 1280 Main St. W, Hamilton, ON L8S 4K1, Canada 
           \and 
Institut de Ci\`encies de l'Espai (ICE), CSIC, Campus UAB, Carrer de Can Magrans s/n, E-08193, Bellaterra, Barcelona, Spain
           \and 
 Institut d'Estudis Espacials de Catalunya (IEEC), E-08860, Castelldefels, Barcelona, Spain
           \and
 Zentrum f\"{u}r Astronomie der Universit\"{a}t Heidelberg,
Institut f\"{u}r Theoretische Astrophysik, Albert-Ueberle-Str. 2, 69120 Heidelberg, Germany
           \and 
 Centre for Astrophysics and Planetary Science, University of Kent, Canterbury, CT2\,7NH, UK
           \and
 Universidad Aut\'onoma de Chile, Nucleo Astroquimica y Astrofisica, Avda Pedro de Valdivia 425, Providencia, Santiago de Chile, Chile }

 
  \abstract
   {Magnetohydrodynamic (MHD) disk winds (DWs) can be responsible for removing angular momentum from the accretion disks of protostars, for driving mass accretion and disk evolution and dispersal, and, ultimately, for setting the conditions for planet formation.}
   {We want to investigate two massive young stellar objects (YSOs), \tari\ and \targ, where evidence for MHD DWs has been previously obtained at scales of 10--100~au through measurements of the 22~GHz water maser velocity distribution within the Protostellar Outflows at the EarliesT Stages (POETS) survey. The different mass, 5.6$\pm$2~\ms\ and $\approx$20~\ms, of the two YSOs allows us to investigate the dependence of MHD DWs on the YSO environment, too.} 
   {We employed IRAM Northern Extended Millimeter Array and archival Atacama Large Millimeter Array observations of \tari\ and \targ, respectively, to study the gas kinematics and physical conditions of the corresponding protostellar winds on scales of 100--1000~au using the same molecular tracers.}
   {In \tari, the emissions of several molecules, particularly SO, SO$_2$, CH$_3$CN, and CH$_3$OH, are distributed along the axis of the previously identified radio jet, and present a local standard of rest (LSR) velocity (\Vlsr) gradient transversal to the jet axis. Position-velocity (PV) plots of the SO lines show patterns consistent with Keplerian rotation. The SO$_2$ emission comes from high-velocity gas flowing close to the jet axis, while CH$_3$CN and CH$_3$OH present a larger radial extension than the S-bearing species. In \targ, the same molecules are instead distributed along the (sky-projected) major axis of the rotating disk reported previously, and their \Vlsr\ gradients consistently trace the disk rotation. The corresponding PV plots present Keplerian profiles. SO is the only molecular species whose emission extends well outside the disk. Our analysis shows that, in both YSOs, the spatial and velocity distributions of SO are consistent with a rotating wind magneto-centrifugally launched from the YSO disk. The comparison with models of molecule formation and excitation in shocks indicates that the different radial extension of the molecular species  observed in the protostellar wind of \tari, as well as the lack of molecules, except SO, in the \targ's wind, can be well explained in terms of a radially extended MHD DW, rather than a compact X-wind.}
   {This study provides compelling evidence that the SO and SO$_2$ emissions can be used to trace MHD DWs in luminous YSOs at scales of 100--1000~au, and confirms the results derived with water masers at scales of 10--100~au.}


   \maketitle
 \nolinenumbers

\section{Introduction}
\label{Int}

The removal of angular momentum is key to allow the surrounding gas to accrete onto the forming star. In an accretion disk, the transport of angular momentum can be driven either by turbulent \citep{Sha73,Lyn74} or magnetic \citep{Pel92,Kon00} stresses. Recent advances in theory \citep{Pud25} and observations    \citep{Mos22,Pas23,DeS24,Nar26} favor models of magnetohydrodynamic (MHD) disk winds \citep[DWs;][]{Bla82,Pud07}, since they can supersede magnetic turbulence and extract disk angular momentum in regions of low ionization, and also account for the commonly observed protostellar outflows, in the form of both slow and poorly collimated winds and fast jets.  In this regard, the magneto-centrifugal acceleration of the wind can take place either at the interface of the stellar magnetosphere with the disk \cite[X-winds;][]{Shu94}, or from the whole centrifugally supported part of the disk (DW). 
While X-winds remove angular momentum only from the innermost disk radii, DWs remove angular momentum across an extended portion of the disk, and therefore they are able to extract from the disk a much larger amount of angular momentum.
This distinction between X-wind and DW is essential to understand the wind's role in driving mass accretion and disk evolution and dispersal, and, ultimately, setting the conditions for planet formation \citep{Kad25}. 

By measuring three-dimensional (3D) velocities and achieving linear resolutions of $\lesssim$1~au, very long baseline interferometry (VLBI) observations of 22~GHz water masers can directly trace the velocity field of protostellar winds \citep{Mos07,San10b,Mos11a}. 
The Protostellar Outflows at the EarliesT Stages \citep[POETS;][]{Mos16,San18,Mos19b} survey has studied the launching region of protostellar winds in a sample ($\approx$~40) of luminous YSOs at linear resolutions of 1--100~au by means of VLBI water maser and Jansky Very Large Array (JVLA) continuum observations. 

One of the most interesting results from POETS is that the distribution of the 3D maser velocities within a few 100~au from the YSOs can be generally interpreted in terms of MHD DWs. There are two kinds of prevalent maser kinematic configurations: (1)~water maser positions and 3D velocities are collimated and trace a fast rotating jet, as in \tari\ \citep[a.k.a. G092.69$+$3.08,][see their Fig.~8]{Mos16}; and (2)~the local standard of rest (LSR) velocity (\Vlsr) gradient of the water masers agrees with that of the underlying rotating disk (observed in ALMA thermal lines) and the proper motions point away from the disk midplane at large angles, as in G011.92$-$0.61 and \targ\ \citep[][see their Figs.~1~and~2, respectively]{Mos19b}. In (1) the water masers trace the fastest and most collimated portion of the MHD DW only (i.e., the jet); in (2) they mainly trace the slower and less collimated DW. If the disk is seen sufficiently close to edge-on, the \Vlsr\ would reflect mainly the disk rotation, while the proper motions would be dominated by the poloidal velocities of the wind. 

It is fundamental to test the MHD DW interpretation based on water maser VLBI at scales of 10--100~au  through interferometric observations of thermal tracers at scales of 100--1000~au. 
Toward G011.92$-$0.61, the case of a MHD DW has been fully confirmed by our Atacama Large Millimeter Array (ALMA) 1.3~mm observations at a linear resolution of $\approx$100~au \citep{Bay25}. The dust and molecular line (e.g., CH$_3$CN and CH$_3$OH) emissions of the rotating disk are well resolved and a strong collimated outflow is traced in SO and SO$_2$ lines up to a distance of $\approx$3400~au. The SO and SO$_2$ emissions show a rotation-dominated velocity pattern, a constant specific angular momentum, and a Keplerian profile that is consistent with a magneto-centrifugal DW origin with launching radii of \ 50--100~au. 

The goal of this article is to further test the water maser MHD DW predictions focusing on the YSOs \tari\ and \targ, representative of the two characteristic maser kinematic configurations observed in POETS. The main properties of the two YSOs are listed in Table~\ref{Tabtar}.
The presence of a rotating molecular disk has been established in both \tari, using IRAM Northern Extended Millimeter Array (NOEMA) observations at a linear resolution of $\approx$600~au \citep{Mos21}, and \targ, employing ALMA Cycle~0 data at a linear resolution of $\approx$1000~au \citep{Bel14}. While \tari\ emits a compact ($\lesssim$~1000~au) radio jet powering a collimated molecular outflow (traced with SO emission at scales from 10$^3$ to 10$^4$~au), \targ\ is associated with an extended nonthermal jet, marked by a collimated chain of radio knots up to distances of 10$^4$~au \citep{San19b}. 

Although the previous millimeter interferometric observations of \tari\ and \targ\ revealed the YSO disk-jet system, the linear resolution was not sufficient to constrain the wind launching mechanism. \tari\ has been recently reobserved with NOEMA at 1.3~mm using long baselines within the large program CORE \citep[P.I.: H.~Beuther,][]{Beu18,Gie21,Ahm23}; for \targ, ALMA Cycle~4 Band~6 archival observations are available. These two interferometric datasets overlap in frequency, and provide comparable linear ($\approx$250~au at the target distance) and velocity ($\approx$0.5~\kms) resolutions, which is optimal for a comparison of the gas kinematics between the two sources. The diverse wind kinematics observed with the water masers at scales of 10--100~au could reflect the different mass (and environment) of the two YSOs (see Table~\ref{Tabtar}). It is therefore particularly interesting to also compare the wind properties at larger scales, to learn how they change with the YSO mass (and environment). 

The structure of the article is as follows. Section~\ref{Obs} describes the NOEMA and ALMA observations. Section~\ref{Res} presents the images of the structure and kinematics of the molecular disk and/or protostellar wind, and determines the gas physical conditions.  In Sect.~\ref{Dis} the launching mechanism of the protostellar winds is discussed, highlighting similarities and differences between the two YSOs. 
Finally, our conclusions are presented in Sect.~\ref{Con}.

\begin{table}[t]
 \caption[]{\label{Tabtar} Properties of the studied YSOs.}
\begin{tabular}{ccccc}
 \hline \hline
 Name & Distance & Luminosity & Mass & Ref. \\
      &  (kpc)   &  ($L_{\odot}$) & ($M_{\odot}$) &  \\
 \hline
 \tari &  1.63$\pm$0.05 & 5$\, \times \,$10$^3$ & 5.6$\pm$2 & 1, 2, 3 \\
 \targ &  2.33$\pm$0.22 & 1--3$\, \times \,$10$^4$ & $\approx$20 & 4, 5, 6 \\
 \hline
\end{tabular}
\tablebib{(1)~\citet{Xu13}; (2)~\citet{Mos16}; (3)~\citet{Mos21};  (4)~\citet{Wu14}; (5)~\citet{Bel14}; (6)~\citet{Olg26}.
}
\end{table}


\begin{table}
\caption{Molecular lines employed in the analysis.}             
\label{lin}      
\centering          
\begin{tabular}{c c c c}     
\hline\hline       
Frequency & Molecule & Transition & $E_{\rm u} / k_{\rm B}$ \\ 
 (GHz)    &          &            &   (K)          \\
\hline 
214.35704  & SO &  7$_8$--7$_7$ & 81 \\
215.22065  & SO &  5$_5$--4$_4$ & 44 \\
219.94944  & SO &  6$_5$--5$_4$ & 35 \\
236.45229  & SO &  1$_2$--2$_1$ & 16 \\
\hline
214.68938  & SO$_2$ & 16$_{3,13}$--16$_{2,14}$ & 148 \\
214.72829  & SO$_2$ & 17$_{6,12}$--18$_{5,13}$ & 229 \\
216.64330  & SO$_2$ & 22$_{2,20}$--22$_{1,21}$ & 248 \\
219.27598  & SO$_2$ & 22$_{7,15}$--23$_{6,18}$ & 353 \\
234.18705  & SO$_2$ & 28$_{3,25}$--28$_{2,26}$ & 403 \\
234.42159  & SO$_2$ & 16$_{6,10}$--17$_{5,13}$ & 213 \\
235.15172  & SO$_2$ & 4$_{2,2}$--3$_{1,3}$ & 19 \\
236.21669  & SO$_2$ & 16$_{1,15}$--15$_{2,14}$ & 131 \\
237.06883  & SO$_2$ & 12$_{3,9}$--12$_{2,10}$ & 94 \\
\hline
220.40390  & CH$_3$CN & 12$_9$--11$_9$ & 647 \\
220.47581  & CH$_3$CN & 12$_8$--11$_8$ & 526 \\
220.53932  & CH$_3$CN & 12$_7$--11$_7$ & 419 \\
220.59442  & CH$_3$CN & 12$_6$--11$_6$ & 326 \\
220.64108  & CH$_3$CN & 12$_5$--11$_5$ & 247 \\
220.67929  & CH$_3$CN & 12$_4$--11$_4$ & 183 \\
220.70902  & CH$_3$CN & 12$_3$--11$_3$ & 133 \\
220.73026  & CH$_3$CN & 12$_2$--11$_2$ & 97 \\
220.74301  & CH$_3$CN & 12$_1$--11$_1$ & 76 \\
220.74726  & CH$_3$CN & 12$_0$--11$_0$ & 69 \\
\hline
217.88650  & CH$_3$OH & 20$_{-1,19}$--20$_{0,20}$ E,  $\varv$$_t$ = 0   & 508 \\
218.44006  & CH$_3$OH & 4$_{-2,3}$--3$_{-1,2}$ E,  $\varv$$_t$ = 0   & 45 \\
220.07856  & CH$_3$OH & 8$_{0,8}$--7$_{-1,6}$ E,  $\varv$$_t$ = 0   & 97 \\
221.29453  & CH$_3$OH & 3$_{2,1}$--2$_{-2,1}$ E,  $\varv$$_t$ = 0   & 40 \\
229.58906  & CH$_3$OH & 15$_{-4,11}$--16$_{-3,14}$ E,  $\varv$$_t$ = 0   & 374 \\
229.86412  & CH$_3$OH & 19$_{5,15}$--20$_{4,16}$ A,  $\varv$$_t$ = 0   & 579 \\
230.02705  & CH$_3$OH & 3$_{2,1}$--4$_{1,4}$ E,  $\varv$$_t$ = 0   & 40 \\
231.28111  & CH$_3$OH & 10$_{2,9}$--9$_{3,6}$ A,  $\varv$$_t$ = 0   & 165 \\
232.41852  & CH$_3$OH & 10$_{2,8}$--9$_{3,7}$ A,  $\varv$$_t$ = 0   & 165 \\
232.78345  & CH$_3$OH & 18$_{3,16}$--17$_{4,13}$ A,  $\varv$$_t$ = 0   & 447 \\
233.79567  & CH$_3$OH & 18$_{3,15}$--17$_{4,14}$ A,  $\varv$$_t$ = 0   & 447 \\
234.68337  & CH$_3$OH & 4$_{2,3}$--5$_{1,4}$ A,  $\varv$$_t$ = 0   & 61 \\
236.93609  & CH$_3$OH & 14$_{1,13}$--13$_{2,12}$ A,  $\varv$$_t$ = 0   & 260 \\
\hline
218.32479  & HC$_3$N   & 24--23  & 131   \\
236512.79  & HC$_3$N   & 26--25  & 153   \\
\hline
217.10492  & SiO &  5--4 & 31 \\
218.22219  & H$_2$CO   &  3$_{0,3}$--2$_{0,2}$  &  21 \\          
219.56036  & C$^{18}$O & 2--1 & 16 \\
220.17742  & CH$_2$CO & 11$_{1,11}$--10$_{1,10}$  &  76 \\
220.39868  & $^{13}$CO & 2--1 & 16 \\
\hline                  
\end{tabular}
\tablefoot{
The molecular data are taken from the Cologne Database for Molecular Spectroscopy \citep[\textsc{CDMS};][]{Mue01,End16}. 
Column~1 reports the rest frequency of the transition; column~2 the molecular species; column~3 the quantum numbers; column~4 the upper state energy.
The quantum numbers are given depending on the symmetry of
the molecule: $J^{\rm upper} - J^{\rm lower}$ for linear, 
$J^{\rm upper}_K - J^{\rm lower}_K$ for symmetric top, 
and $J^{\rm upper}_{K_a,K_c} - J^{\rm lower}_{K_a,K_c}$ for asymmetric top molecules.
}
\end{table}


\begin{table}
\caption{Physical parameters from the fit of the molecular lines.}             
\label{mol_fit}      
\centering      
\small    
\begin{tabular}{c c c c c c}     
\hline\hline       
Species & $T_{\rm ex} $ & $N_{\rm col}$  & $\varv_0$   & FWHM  & $\langle\tau\rangle$, $\tau_{\rm max}$ \\ 
     &   (K)     &   ($10^{16}$ cm$^{-2}$) &   (\kms) &   (\kms)    & \\
\hline 
  \multicolumn{6}{c}{IRAS~21078$+$5211 YSO} \\
\hline
SO   & 33$\pm$3 & 1.3$\pm$0.4 & $-$7.0$\pm$0.3 & 7.6$\pm$0.8 &  0.2, 5 \\
SO$_2$ & 168$\pm$13 & 4.7$\pm$0.3 & $-$6.8$\pm$0.2 & 9.0$\pm$0.5 &  0.08, 0.18 \\
CH$_3$OH & 217$\pm$7 & 43.4$\pm$1.6 & $-$6.5$\pm$0.1 & 7.8$\pm$0.2 & 0.07, 0.23 \\
CH$_3$CN & 320$\pm$50 & 1.0$\pm$0.2 & $-$6.9$\pm$0.2 & 9.3$\pm$0.5 & 0.05, 0.11 \\
HC$_3$N & 78$\pm$26 & 0.07$\pm$0.02 & $-$6.3$\pm$0.3 & 10.8$\pm$0.7 & 0.17, 0.18 \\
\hline 
  \multicolumn{6}{c}{IRAS~21078$+$5211 Outflow} \\
  \hline
\multicolumn{6}{c}{along the axis:} \\
SO  & 30$\pm$2 & 1.8$\pm$2.5 & $-$7.0$\pm$0.2 & 3.6$\pm$0.8 & 0.1, 16 \\
SO$_2$ & 132$\pm$25 & 1.6$\pm$0.2 & $-$9.1$\pm$0.5 & 8.3$\pm$1.2 & 0.04, 0.10 \\
CH$_3$OH & 207$\pm$10 & 26.2$\pm$1.6 & $-$8.6$\pm$0.1 & 5.6$\pm$0.3 & 0.03, 0.22 \\
CH$_3$CN & 165$\pm$29 & 0.19$\pm$0.05 & $-$8.1$\pm$0.2 & 4.7$\pm$0.4 & 0.09, 0.21\\
HC$_3$N & 43$\pm$29 & 0.05$\pm$0.07 & $-$9.5$\pm$0.5 & 5.9$\pm$1.3 & 0.17, 0.20 \\
\multicolumn{6}{c}{far from the axis:} \\
SO  & 18$\pm$2 & 1.2$\pm$0.9 & $-$5.3$\pm$0.3 & 4.0$\pm$0.7 & 7.3, 15  \\
SO$_2$ & 118$\pm$9 & 0.78$\pm$0.05 & $-$6.4$\pm$0.1 & 4.4$\pm$0.3 & 0.04, 0.12 \\
CH$_3$OH & 216$\pm$7 & 34.8$\pm$1.4 & $-$5.5$\pm$0.1 & 4.5$\pm$0.1 & 0.10, 0.33 \\
CH$_3$CN & 133$\pm$15 & 0.16$\pm$0.01 & $-$5.5$\pm$0.1 & 3.8$\pm$0.2 & 0.13, 0.32\\
HC$_3$N & 58$\pm$57 & 0.02$\pm$0.02 & $-$5.7$\pm$0.6 & 6.0$\pm$1.5 & 0.07, 0.08 \\ 
\hline 
  \multicolumn{6}{c}{G035.02$+$0.35 YSO} \\
\hline
SO$_2$ & 145$\pm$5 & 12.5$\pm$0.6 & 49.6$\pm$0.3 & 12.2$\pm$0.6 & 0.11, 0.20 \\
CH$_3$OH & 257$\pm$10 & 68.9$\pm$4.1 & 45.4$\pm$0.1 & 5.7$\pm$0.2 & 0.11, 0.25 \\
CH$_3$CN & 470$\pm$56 & 2.1$\pm$0.2 & 46.6$\pm$0.1 & 6.7$\pm$0.3 &  0.06, 0.11 \\
\hline                  
\end{tabular}
\tablefoot{\\
Results of the \textsc{MADCUBA} fits of the spectra extracted toward the YSO for both \tari\ and \targ\, and toward the outflow for \tari\ only (see Sect.~\ref{Phy}). Column~1 reports the molecular species; columns~2,~3,~4,~and~5 give, respectively, the excitation temperature, the column density, the velocity, and the line width of the fit molecular lines; column~6 lists the average and maximum optical depth of the fit molecular lines.
}
\end{table}

\section{Observations}
\label{Obs}

\subsection{NOEMA observations}
\label{Obs_N}
\tari\ (tracking center: RA(J2000) = $21^{\rm h} \, 09^{\rm m}$ 21\pss64, Dec(J2000) = $+52$\degree\  $22^{\prime}$ 37\pas5) was observed with the new long baselines in the new A configuration at NOEMA during the winter term 2022--2023 in three tracks between February~25 and March~4, 2023, in project W22AL003 (P.I.: H.~Beuther). The covered baseline range was between $\approx$58 and $\approx$1690~m. The quasars 2013$+$370 and MWC349 were employed for bandpass and flux calibration, respectively. Gain calibration was performed via regularly interleaved observations of the quasars 2037$+$511 and 2022$+$542. The spectral coverage in the 1.3~mm band in the lower and upper sidebands was between $\approx$213.85 to $\approx$221.60~GHz and $\approx$229.37 to $\approx$237.17~GHz, respectively. The spectral resolution of 0.25~MHz corresponds at 220~GHz to a nominal velocity resolution of $\approx$0.34~\kms. In our data reduction we employed a uniform spectral resolution of 0.5~\kms. The assumed nominal \Vlsr\ of the region is $-$6.1~\kms.

Calibration and imaging of the data was conducted within the {\textsc GILDAS} framework\footnote{http://www.iram.fr/IRAMFR/GILDAS} with {\textsc clic} and {\textsc mapping}. By excluding strong line emission and combining the lower and upper sideband data, we obtained a double-sideband (DSB) continuum dataset for further imaging. The continuum data were self-calibrated within {\textsc mapping} in phase-only in three loops with decreasing integration times of 300, 150 and 45~s. Imaging was then conducted in uniform weighting to achieve the highest spatial resolution. The final DSB continuum images have a synthesized beam with full width at half maximum (FWHM) major and minor sizes of \ 0\farcs193 $\times$ 0\farcs103 \ and a position angle (PA) of $-$173\degree. The continuum $1\sigma$ root mean square (rms) noise is 0.09~m\Jyb.

While the self-calibration improved the continuum data significantly with an increase in the signal-to-noise ratio of the maps from 19 to 136, no clear improvement of the line data was recognizable after applying the gain solutions from the continuum phase self-calibration. Therefore, for the spectral lines, we used the non-self-calibrated data products. The rms noise in 0.5~\kms\ channels is typically a few milliJansky per beam (or a few Kelvin in brightness temperature), with a synthesized beam comparable to the continuum data as described above. 

\subsection{ALMA archive data}
\label{Obs_A}
\targ\ (tracking center: RA(J2000) = $18^{\rm h} \, 54^{\rm m}$ 00\pss6485, Dec(J2000) = $+02$\degree\  $01^{\prime}$ 19\pas278) was observed within the project 2016.1.01036.S (P.I.: P.~Sanhueza) by ALMA in Cycle~4 on September 12, 2017, employing the extended configuration (C40-8). The phase calibrator was the quasar J1851$+$0035. 43 antennae of the 12~m Array took part to the observations covering baselines from \ $\approx$92~m to $\approx$8548~m. Four spectral windows (SPWs) that had a width of 1875~MHz centered at the sky frequency of 217.812, 220.013, 232.013, and 234.513~GHz were recorded. The correlator spectral channels were 1920 for the SPWs centered at 232.013 and 234.513~GHz, and 3840 for the SPWs centered at 217.812 and 220.013~GHz, providing a maximum velocity resolution of  0.67~\kms.

The restoring calibration employed the ALMA pipeline version 42254 of the Common Astronomy Software Applications \citep[\textsc{CASA};][]{CASA22} package, version 5.4.0-70. The image of each SPW (continuum plus line emission) was produced manually using the \textsc{tclean} task, using Hogbom deconvolution \citep{Hog74} and Briggs weighting \citep{Bri95}, with the robust parameter set to 0.5 as a compromise between resolution and sensitivity to extended emission. The clean beams of the resulting images have FWHM  major and minor sizes varying in the range  \ 0\farcs12--0\farcs14 and 0\farcs09--0\farcs11, and PA in the range [$-$67\degree, $-$79\degree]. 

For each SPW, to determine the continuum level of the spectra and subtract it from the line emission, we used \textsc{STATCONT} \citep{Sanc18}, a statistical method of estimating the continuum level at each position of the map from the spectral distribution of the intensity at that position.
Taking the average of the four SPWs, the 1.3~mm continuum image of \targ\ has a 1$\sigma$ rms noise level of 0.32~m\Jyb. The 1$\sigma$ rms noise in a single spectral channel varies in the interval 3--4.5~m\Jyb\ (or 5--8~K in brightness temperature), depending on the considered SPW.


  \begin{figure*}
    \includegraphics[width=0.49\textwidth]{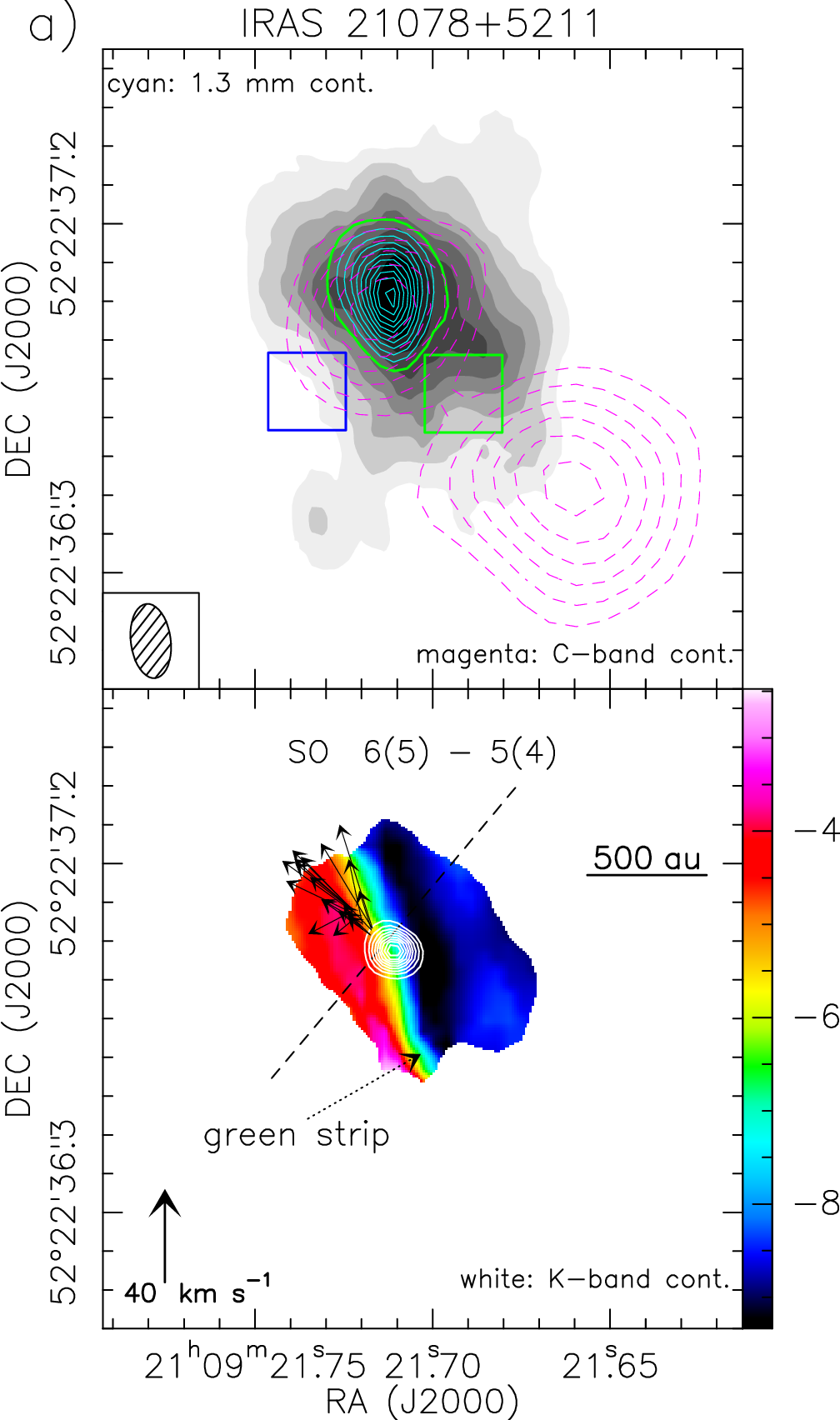} 
    \includegraphics[width=0.515\textwidth]{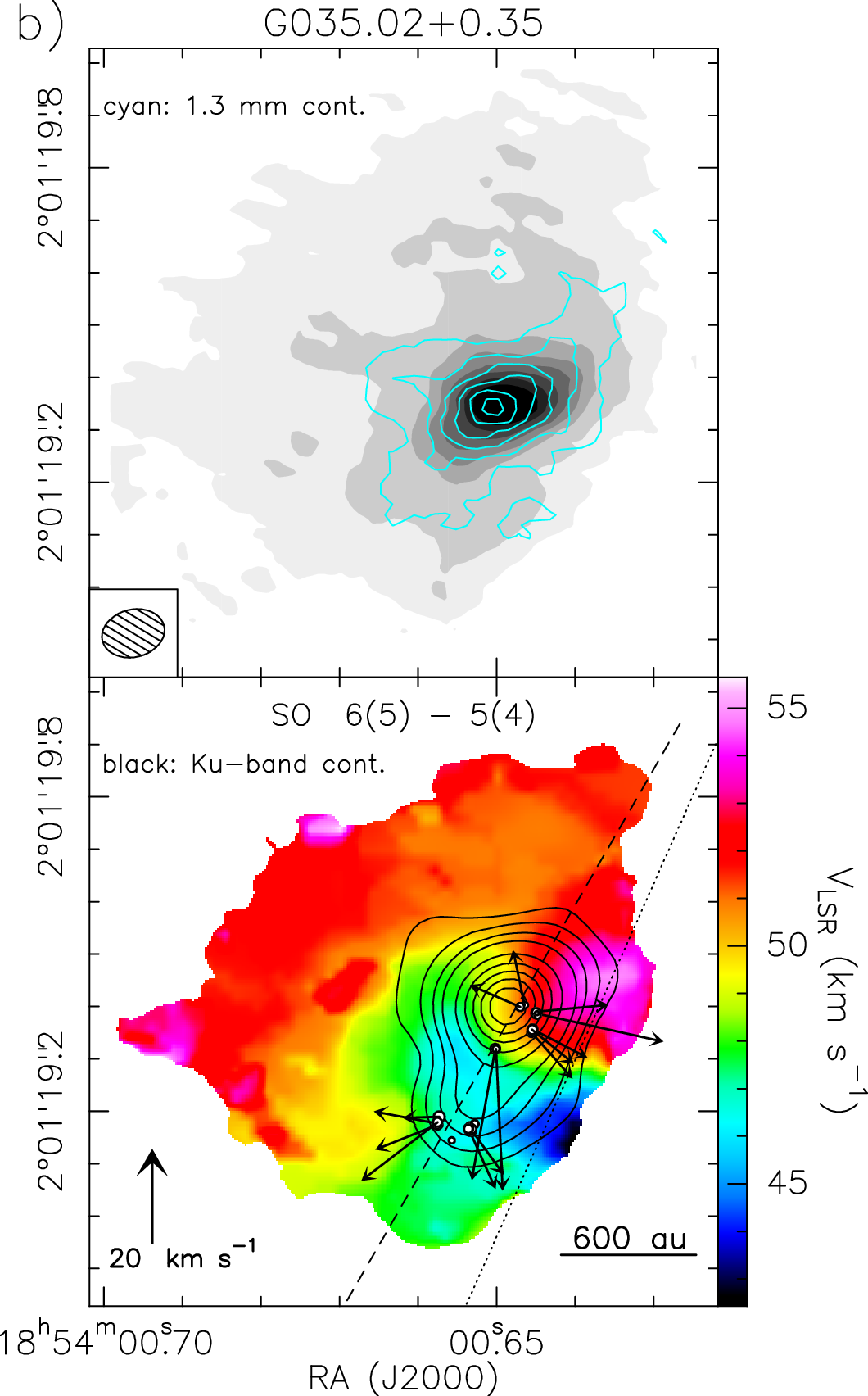}
    \caption{Rotating wind in \tari\ (panel~a) and \targ\ (panel~b). In both panels, the grayscale and color maps in the upper and lower plots correspond to the velocity-integrated emission and intensity-weighted velocity, respectively, of the SO 6$_5$--5$_4$ line. The green color represents \Vlsr\ close to the systemic velocity. \ (Panel~a)~The level contours of the grayscale filled-contours map in the upper plot are from 10\% to 80\% in steps of 10\% of the map peak of 0.25~\Jyb~\kms. Cyan contours reproduce the 1.3~mm continuum showing levels from 1.9 to 11.9~m\Jyb\ at steps of 1.0~m\Jyb. The JVLA C-~and~K-band continuum emissions are given by magenta and white contours, respectively, and black arrows show the proper motions of the 22~GHz water masers \citep[][see their Fig.~8]{Mos16}. The dashed black line marks the direction of the disk midplane at PA$\approx$140\degree\ (see Sect.~\ref{Stru}). The thick green contour and thick green and blue squares in the upper~plot delimitate the integration area for the spectra toward the YSO and outflow (see Sect.~\ref{Phy}). In the lower plot, the green strip (see Sects.~\ref{Stru}~and~\ref{McwI}) is indicated. \ (Panel~b)~The level contours of the grayscale filled-contours map in the upper plot are 5\%, 15\%, and from 30\% to 80\% in steps of 10\% of the map peak of 0.70~\Jyb~\kms. Cyan contours reproduce the 1.3~mm continuum showing levels from 1.2 to 6.2~m\Jyb\ at steps of 1.0~m\Jyb. The JVLA Ku-band continuum emission is shown with black contours, and black-edge white dots and black arrows give the positions and proper motions of the 22~GHz water masers \citep[][see their Fig.~2]{Mos19b}. The dashed black line marks the direction of the disk midplane at PA$\approx$150\degree\ (see Sect.~\ref{Stru}). The dotted black line crossing the loci of maximum and minimum \Vlsr\ indicates the approximate direction of the SO \Vlsr\ gradient.}
    \label{SO}
   \end{figure*}


  \begin{figure*}
  \centering
    \includegraphics[width=0.845\textwidth]{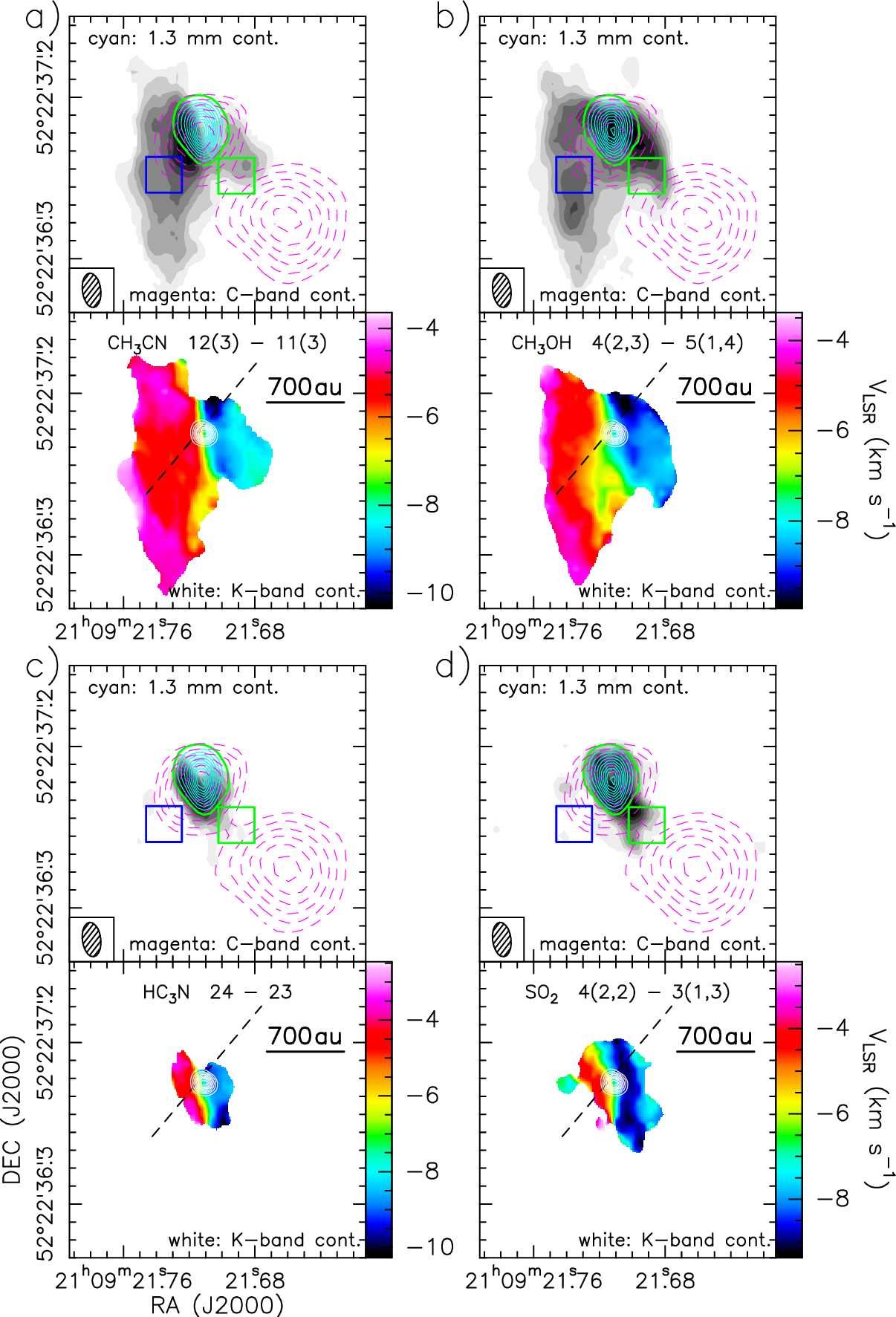} 
    \caption{Structure and kinematics of the \tari\ wind in molecular tracers. Each of the four panels presents the maps of the velocity-integrated emission (upper~plot) and intensity-weighted velocity (lower~plot) for a specific molecular line, as labeled. The level contours of the grayscale filled-contours maps in the upper plots are from 10\% to 80\% in steps of 10\% of the map peak. Contours, lines, and symbols have the same meanings as in Fig.~\ref{SO}a.}
    \label{I21_mol}
   \end{figure*}


  \begin{figure*}
  \centering
   \includegraphics[width=0.835\textwidth]{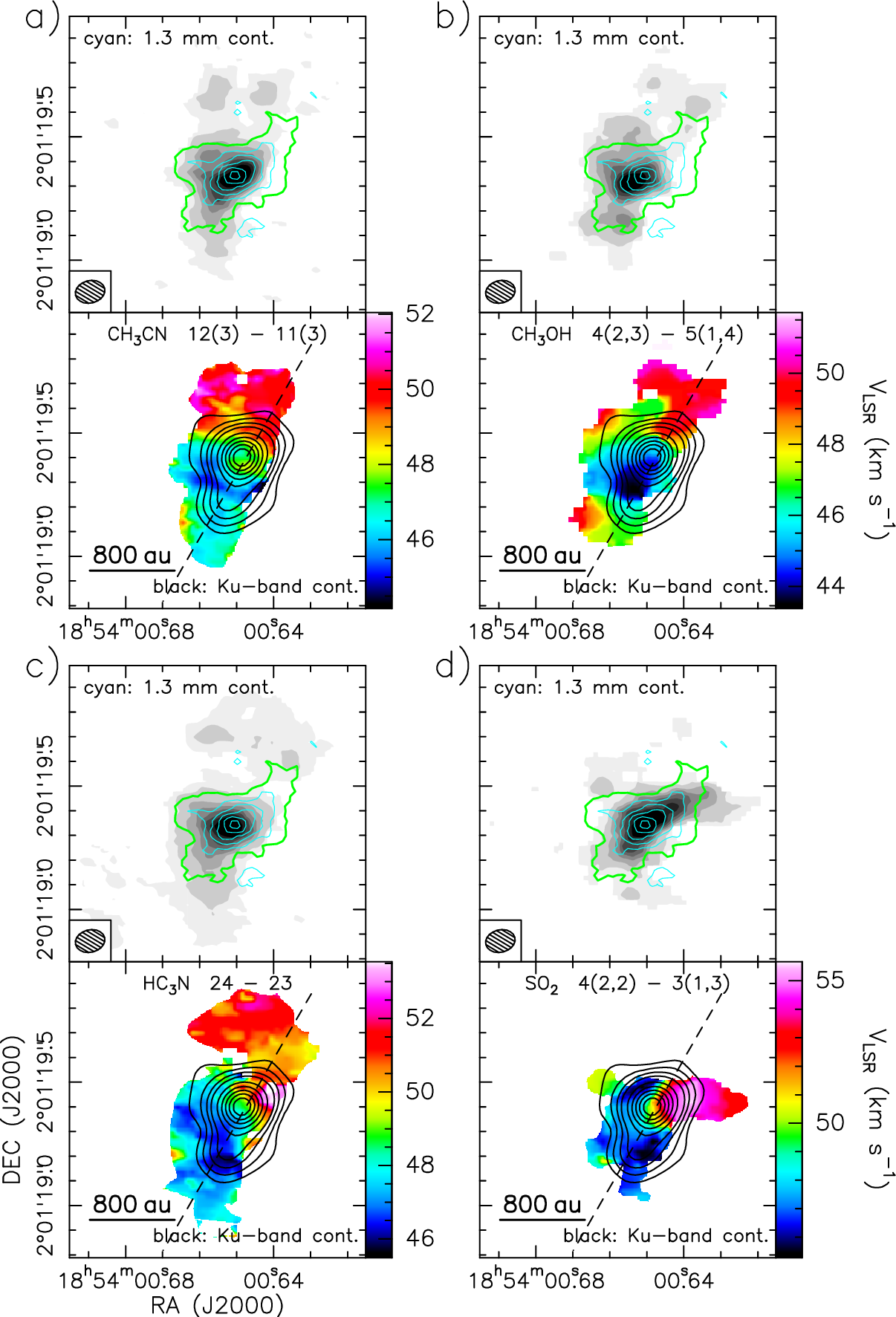} 
    \caption{Structure and kinematics of the \targ\ disk in molecular tracers. Each of the four panels presents the maps of the velocity-integrated emission (upper~plot) and intensity-weighted velocity (lower~plot) for a specific molecular line, as labeled. The level contours of the grayscale filled-contours maps in the upper plots are 5\%, 15\%, and from 30\% to 80\% in steps of 10\% of the map peak. The thick green contour in the upper~plots delimitates the integration area for the spectrum toward the YSO (see Sect.~\ref{Phy}). Other contours and lines have the same meanings as in Fig.~\ref{SO}b.}
    \label{G035_mol}
   \end{figure*}


\section{Results}
\label{Res}

\subsection{Structure and kinematics}
\label{Stru}

Table~\ref{lin} lists the main molecular lines we considered to determine the gas kinematics and physical conditions. Most of them are covered by both the NOEMA \tari\ and the ALMA \targ\ observational setups, which allows us to investigate the two sources by employing the same molecular tracers. To study the kinematics, we surveyed the spectral images and selected the molecular lines that satisfy two conditions: \ 1)~being sufficiently intense to map an extended velocity range; \ 2)~being representative of well-defined kinematic structures. Our choice includes typical tracers of massive disks, such as CH$_3$CN and CH$_3$OH \citep{Ces99, Bel04,Mos19,San19a}, and protostellar winds in low-~and~high-mass YSOs, such as SO and SO$_2$ \citep{Tab17,Bay25}. For completeness, we note that the SiO 5--4 emission, another potential outflow tracer, is resolved out at these small scales.

Figure~\ref{SO} presents the structure and kinematics of the gas traced by the SO 6$_5$--5$_4$ line in both targets. In \tari\ the SO emission extends over $\approx$1500~au in the jet direction at PA $\approx$ 50\degree\ \citep{Mos24}, marked by both the double-lobe JVLA C-band emission and the collimated proper motions of the water masers, and presents a well-defined \Vlsr\ gradient transversal to the jet axis and parallel to the disk midplane (the dashed black line in Fig.~\ref{SO}a, lower plot). The \Vlsr\ pattern of the SO emission in \targ\ is more complex, but the loci of maximum and minimum \Vlsr\ define a direction for the \Vlsr\ gradient (marked by the dotted black line in Fig.~\ref{SO}b, lower plot) that is roughly parallel to the (sky-projected) major axis (the dashed black line in Fig.~\ref{SO}b, lower plot) of the rotating molecular disk at PA $\approx$150\degree\ \citep{Bel14}.
The same direction corresponds also to the elongation of the radio continuum and the \Vlsr\ gradient traced by the water masers \citep[][see their Fig.~2]{Mos19b}. It is interesting to note that the SO \Vlsr\ gradient is slightly offset to the southwest with respect to the radio continuum and water maser distribution. While the core of the SO emission is well aligned with the 1.3~mm continuum, more diffuse SO emission extends beyond the 1.3~mm continuum to the northeast and protrudes in spurs oriented in a direction approximately parallel to the disk rotation axis (see Fig.~\ref{SO}b, upper~plot).

Figures~\ref{I21_mol}~and~\ref{G035_mol} compare the spatial distribution and kinematics of dust and gas at scales of 100--1000~au in \tari\ and \targ\ employing the same tracers. The gas distribution is traced with bright lines of CH$_3$CN, CH$_3$OH, HC$_3$N, and SO$_2$. While the 1.3~mm continuum is only slightly resolved in \tari, in \targ, at comparable linear resolution, it presents an extended structure approximately parallel to the disk midplane. In both regions, all the molecular lines trace a well-defined \Vlsr\ gradient along the direction of the disk. Specifically for \tari, a characteristic feature of all the velocity maps, although best visible in the SO emission, is the green strip that bisects the maps, forming a small angle with the jet axis (see Figs.~\ref{SO}~and~\ref{I21_mol}).
The most striking difference between the two YSOs is in regard to the elongation of the molecular emissions, which are parallel to the jet in \tari\ and to the disk in \targ. In both targets, the emission of SO$_2$ is more compact than that of CH$_3$CN and CH$_3$OH, particularly in \tari, where the latter covers a much wider area around the jet axis.  

\subsection{Physical conditions}
\label{Phy}

To determine the physical conditions of the gas, we used the spectral analysis software Madrid Data Cube Analysis \citep[\textsc{MADCUBA}\footnote{https://cab.inta-csic.es/madcuba/};][]{Mar19} to identify and fit the emission of several molecular species. For both \tari\ and \targ, we derived the spectral emission in the direction of the YSO by integrating the signal over the area inside the lowest 1.3~mm continuum contour (the thick green contours in Figs.~\ref{I21_mol}~and~\ref{G035_mol}, respectively). In \tari, since the molecular emission is significantly more extended than the 1.3~mm continuum, we also extracted spectra at two positions in the outflow: \ 1)~along the outflow axis (inside the thick green square in Fig.~\ref{I21_mol} centered at RA(J2000) = $21^{\rm h} \, 09^{\rm m}$ 21\pss691 and Dec(J2000) = $+52$\degree\  $22^{\prime}$ 36\pas76, with size of 0\farcs2); \ 2)~at a wide separation from the axis (inside the thick blue square in Fig.~\ref{I21_mol} centered at RA(J2000) = $21^{\rm h} \, 09^{\rm m}$ 21\pss735 and Dec(J2000) = $+52$\degree\  $22^{\prime}$ 36\pas77, with a size of 0\farcs2). 

Using the \textsc{SLIM} (Spectral Line Identification and Modeling) tool of \textsc{MADCUBA}, we surveyed the spectra to search for molecular species with a relatively large number of unblended and optically thin lines suitable for deriving the gas physical conditions in the inner regions close to the YSO. We selected the CH$_3$CN, CH$_3$OH, and SO$_2$ molecules (see Table~\ref{mol_fit}) that in both targets have optical depths of less (or much less) than\ 0.25, and cover a sufficiently wide range in excitation energy to allow for a reliable estimate of the excitation temperature. In \tari, we fit the emission of SO and HC$_3$N, too, since, owing to the wider frequency coverage of the observational setup,  multiple lines are observed for these molecules, too. 

To derive the physical parameters of molecular gas, we used the tool \textsc{AUTOFIT} of \textsc{MADCUBA}, which compares the observed spectra with the LTE synthetic spectra, taking into account all transitions and the line opacities. We warn that a single component was fit for each molecular species, under the assumption of negligible density and temperature gradients. Leaving four parameters free to vary, that is, the column density ($N_{\rm col}$), excitation temperature ($T_{\rm ex}$), line \Vlsr\ ($\varv_0$), and line width (FWHM), \textsc{AUTOFIT} provides the best nonlinear least-squares fit using the Levenberg-Marquardt algorithm \citep{Lev44,Mar63}. 
Nominal errors of the fit parameters were derived from the diagonal elements of the covariance matrix calculated at the $\chi^2$ minimum.
Figures~\ref{I21_fit_ch3cn-hc3n}, \ref{I21_fit}, and~\ref{G035_fit} present the result of the fit for a subset of the more unblended and intense lines of each molecular species. The emission profile of each transition is reasonably well fit with \textsc{MADCUBA}, determining the values of the column density, temperature, velocity, and line width listed in Table~\ref{mol_fit}.


  \begin{figure}
   \includegraphics[width=\hsize]{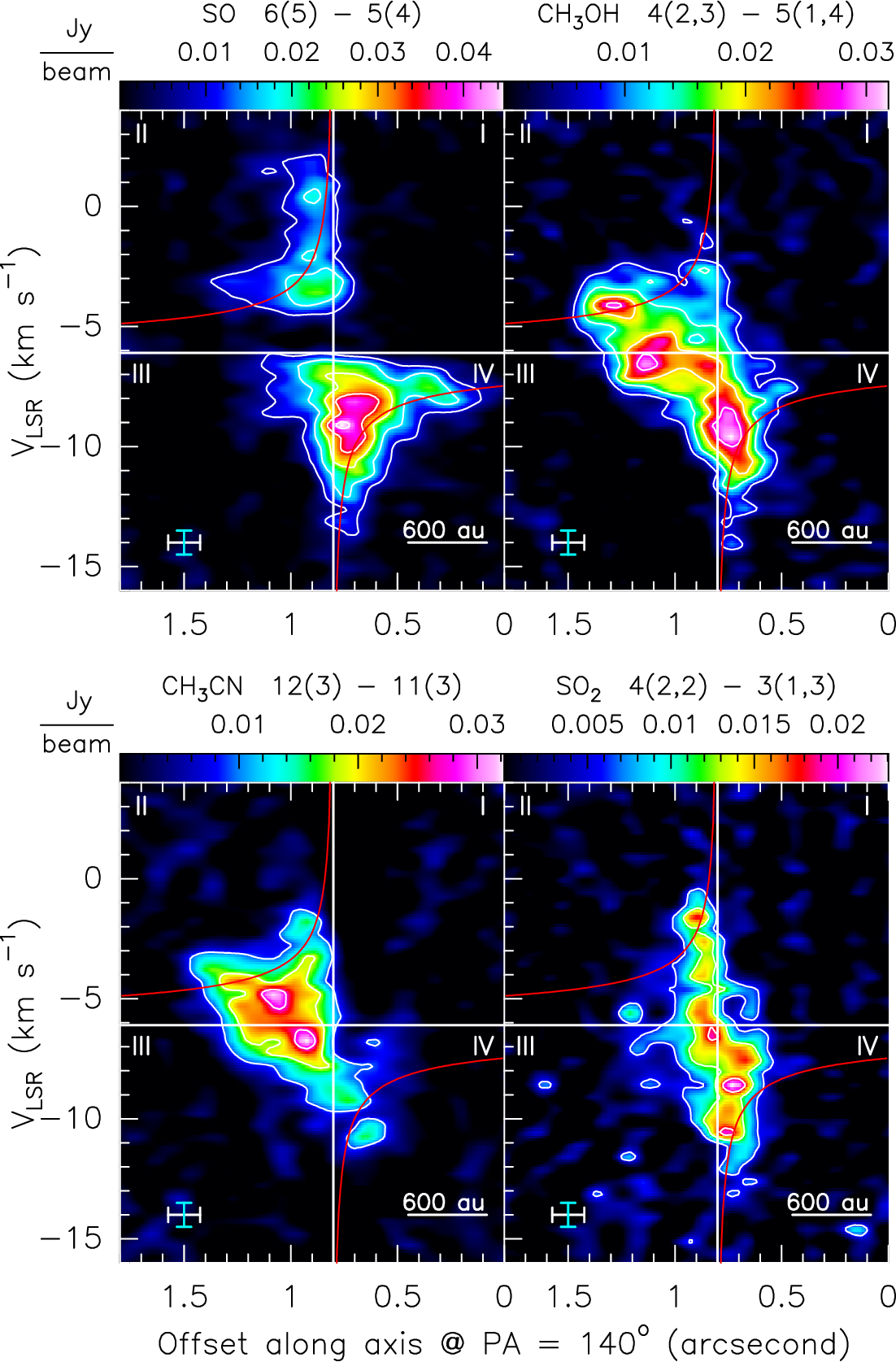} 
    \caption{Gas kinematics in \tari. Each of the four panels presents the PV plot along the cut at PA = 140\degree \ (marked by the dashed black line in Figs.~\ref{SO}~and~\ref{I21_mol}) of the molecular line specified above the panel. PV plots are shown with color maps and white contours, with contour levels ranging from $\left| I_{\rm min} \right|/2$ \ to \ $I_{\rm max}$ \ at steps of \ $\left| I_{\rm min} \right|/2$, where \ $I_{\rm min}$ \ and \  $I_{\rm max}$ \ are the minimum and maximum of the map, respectively. In each panel, the vertical and horizontal white axes denote the positional offset ($\approx$ 0\farcs8) of the YSO relative to the phase center of the observations and YSO \Vlsr\ ($\approx$ $-$6.1~\kms), respectively. The four quadrants are labeled with roman numbers. For comparison, the red curve shows the Keplerian velocity profile for a central mass of 6~\ms. In the lower left corner of the panels, the vertical cyan and horizontal white error bars indicate the velocity and spatial resolutions, respectively. }
    \label{I21_PV}
   \end{figure}


  \begin{figure}
 \includegraphics[width=\hsize]{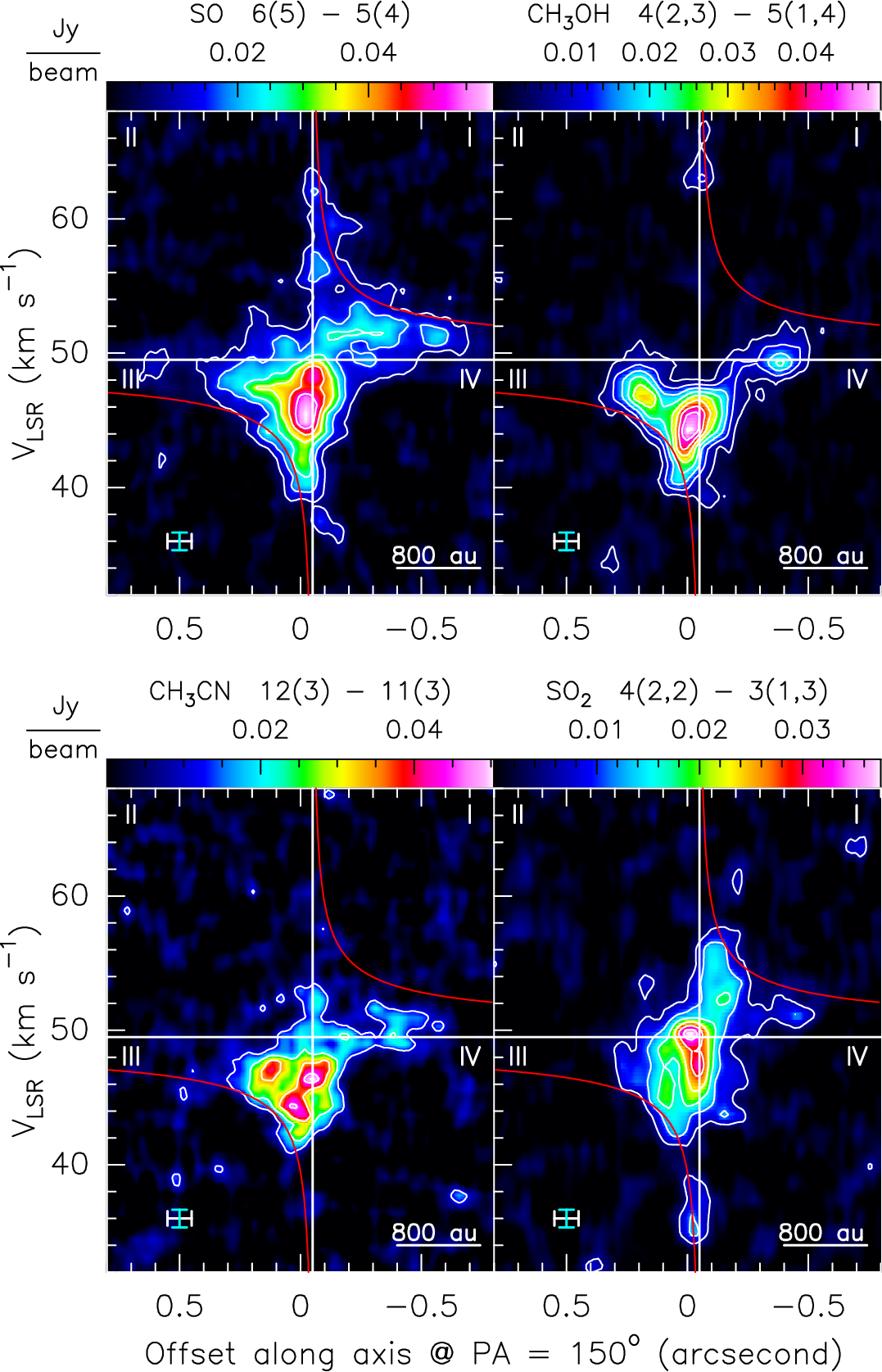} 
    \caption{Gas kinematics in \targ. Each of the four panels presents the PV plot along the cut at PA = 150\degree \ (marked by the dashed black line in Figs.~\ref{SO}~and~\ref{G035_mol}) of the molecular line specified above the panel. PV plots are shown with color maps and white contours, with contour levels ranging from $\left| I_{\rm min} \right|/2$ \ to \ $I_{\rm max}$ \ at steps of \ $\left| I_{\rm min} \right|/2$, where \ $I_{\rm min}$ \ and \  $I_{\rm max}$ \ are the minimum and maximum of the map, respectively. 
In each panel, the vertical and horizontal white axes denote the positional offset ($\approx$ $-$0\farcs05) of the YSO relative to the phase center of the observations and YSO \Vlsr\ ($\approx$ 49.5~\kms), respectively.    
The four quadrants are labeled with roman numbers. For comparison, the red curve shows the Keplerian velocity profile for a central mass of 20~\ms. In the lower left corner of the panels, the vertical cyan and  horizontal white error bars indicate the velocity and spatial resolutions, respectively.}
    \label{G035_PV}
   \end{figure}

\section{Discussion}
\label{Dis}

To study the gas kinematics, we constructed position-velocity (PV) plots of the molecular emissions along the YSO disk midplane (marked by the dashed black line in Figs.~\ref{SO}, \ref{I21_mol},~and~\ref{G035_mol}). Before producing the PV plot, the emission was averaged across a beam in the direction perpendicular to the positional cut. Figures~\ref{I21_PV}~and~\ref{G035_PV} show the PV plots of representative lines of SO, CH$_3$OH, CH$_3$CN, and SO$_2$ for \tari\ and \targ, respectively.

\subsection{Magneto-centrifugal winds}
\label{Mcw}

In a magneto-centrifugal wind, gas particles are launched along a streamline with a specific angular momentum higher than that of the gas in the disk. A general outcome of the models is that, along a given streamline, the rotational velocity of the wind, starting from the Keplerian value at the launch radius, stably grows till reaching the Alfv\'en point, then decreases beyond that point owing to the significant inertia of the flow, and finally increases again due to conservation of angular momentum when the wind starts recollimating \citep[see, e.g.,][]{Tab20}. Based on this qualitative arguments, at distances of $\sim$100~au from the disk midplane of YSOs of $\sim$10~\ms, we can expect that a magneto-centrifugal wind displays a Keplerian rotation profile comparable to that of the gas rotating in the disk.

\subsubsection{\tari}
\label{McwI}

In \tari, all the molecular emissions are elongated in the direction of the jet and trace a common \Vlsr\ gradient transversal to the jet axis. These two findings provide strong evidence for a rotating wind emerging from the YSO. It is best traced by the emissions of SO (see Fig.~\ref{SO}a) and SO$_2$ (see Fig.~\ref{I21_mol}d), which, being typical outflow tracers, present the best collimation along the jet axis. Besides, looking at Fig.~\ref{I21_PV}, the SO velocity profile, which is one of the most extended in radius, shows the best indication of Keplerian rotation. The PV plots in Fig.~\ref{I21_PV} were obtained by averaging the emission perpendicularly to the disk midplane across a distance of $\approx$~250~au much longer than the expected height of the disk atmosphere. If SO is mainly excited in the wind, its velocities should reliably trace the gas motion at the base of the wind. 
The contribution of the disk kinematics on the PV plot of SO in Fig.~\ref{I21_PV} can be evaluated by comparing it with PV plots at positions along the jet axis sufficiently far from the disk midplane (see Fig.~\ref{I21_PV_off}) such that only the outflow motion can contribute. These off-disk PV plots show velocity profiles consistent with that at the disk midplane, confirming that the SO emission traces mainly the wind kinematics.

Comparing the PV plots of SO and SO$_2$, with respect to SO, SO$_2$ is emitted from gas projected on the sky closer to the jet axis, which probably reaches the highest (absolute) velocities at the smallest line-of-sight distances from the YSO. Ultimately, the kinematics traced by SO and SO$_2$ suggests a magneto-centrifugal launch of the wind, where the rotation velocities of the wind streamlines follow a Keplerian pattern similar to that of their launch radii in the disk.

We can reasonably rule out that the observed \Vlsr\ gradient transversal to the jet axis is due to the rotation of the ambient gas induced by the gravitational force of the YSO. Let us indicate with \ $Z$ (increasing to northeast) \ and \ $R$ (increasing to northwest) \ the distances from the YSO measured along and across the jet axis. Assuming centrifugal equilibrium we can write
\begin{equation}
\label{VR}
\frac{V_{\rm rot}^2}{R} = F_{\rm g}(Z,R) = \frac{G \, M}{Z^2 \, + \, R^2} \; \frac{R}{\sqrt{Z^2 \, + \, R^2}}  
,\end{equation}
where $V_{\rm rot}$ is the rotational velocity, $M$ \ is the YSO mass, and $F_{\rm g}(Z,R)$ \ is the gravitational force at the position $(Z, R)$ \ corrected by the factor \ $R \, / \, \sqrt{Z^2 \, + \, R^2}$ \ to take into account its inclination with respect to the rotation radius. From Eq.~\ref{VR}, the derivative of the rotational velocity with $R$ is given by
\begin{equation}
\label{dVR}
\frac{{\rm d }V_{\rm rot}}{{\rm d }R} \approx 0.95 \; \frac{\sqrt{M}}{\sqrt{d}}  \; \frac{Z^2 \, - \, 0.5 \, R^2}{[Z^2 \, + \, R^2]^{7/4}}
,\end{equation}
where the rotation velocity, \ $V_{\rm rot}$, \ is measured in kilometers per second, $Z$ \ and \ $R$ \ in arcseconds, and $M$ \ and $d$ \ are in solar mass and kiloparsecs. Let us consider the positions \ $(Z_{\rm gs}$,$R_{\rm gs})$ \ along the green strip, where, for our choice of velocity-color code, the transversal \Vlsr\ gradient is best visible.
Along this strip we have \ $\left|Z_{\rm gs}\right| > \left|R_{\rm gs}\right|$ \ and \  $\left|Z_{\rm gs}\right|^2 \gg \left|R_{\rm gs}\right|^2$, such that we can approximate Eq.~\ref{dVR} with
\begin{equation}
\label{dVRap}
\frac{{\rm d }V_{\rm rot}}{{\rm d }R} \approx 0.95 \; \frac{\sqrt{M}}{\sqrt{d}}  \; \left|Z_{\rm gs}\right|^{-3/2}
.\end{equation}
Across the green strip one observes \Vlsr\ changes as large as \ $\Delta V_{\rm rot}$~$\approx$~6~\kms\ over radial offsets as small as  \ $\Delta R$~$\approx$~0\farcs2 (see Fig.~\ref{SO}a, lower~plot). Using these values in Eq.~\ref{dVRap} to calculate the derivative \ ${\rm d }V_{\rm rot} / {\rm d }R$, at the extremes of the green strip where  \ $\left|Z_{\rm gs}\right| \approx$ 0\farcs3, we derive a value for the YSO mass, \ $M$~$\approx$~44~\ms. This value greatly exceeds the value of the YSO mass of 5.6$\pm$2~\ms\ determined by studying the kinematics of the molecular lines tracing the disk rotation \citep{Mos21}. Therefore, we can conclude that the observed \Vlsr\ gradient is by far too large to correspond to the rotation of the YSO envelope, and has to be linked instead with the rotation of the wind.

In Sect.~\ref{Stru} we noted the green strip that bisects the velocity maps of \tari. Since the green color represents the YSO \Vlsr\ ($\approx$ $-$6.1~\kms), one can see that the \Vlsr\ along the jet axis is redshifted to northeast and blueshifted to southwest. That agrees with the polarity of the SO outflow observed at larger scales \citep[][see their Fig.~9]{Mos21}. Thus, the straightforward interpretation of the tilt of the green strip with respect to the jet axis is that the jet is being accelerated at scales of $\sim$100~au, such that the \Vlsr\ along the jet axis is progressively more redshifted (blueshifted) to northeast (southwest). The fact that the loci where the jet \Vlsr\ equals that of the YSO are found (approximately) along a line might be explained assuming that the change of \Vlsr\ is (approximately) linear for small displacements. At a given position, \ $Z$, \ along the axis, the jet acceleration causes a \Vlsr\ variation from the systemic velocity \ $\Delta V_{\rm ax} = C_{\rm ax} \; Z$, with $C_{\rm ax}$ being a constant. Similarly, in the radial direction, the rotation produces a change  from the axial \Vlsr\ \ , $\Delta V_{\rm rot} = C_{\rm rot} \; R$. At the positions \ $(Z_{\rm gs}$,$R_{\rm gs})$ \ of the green strip, $\left|\Delta V_{\rm ax}\right| = \left|\Delta V_{\rm rot}\right|$ \ and \ $R_{\rm gs} / Z_{\rm gs} = C_{\rm rot} / C_{\rm ax}$. The constant ratio between \ $R_{\rm gs}$ \ and \ $Z_{\rm gs}$ \ can explain the occurrence of the green strip.

\subsubsection{\targ}
\label{McwG}

In \targ, at scales of $\approx$1000~au, all the molecular lines (see Fig.~\ref{G035_mol}), with the exception of SO (see Fig.~\ref{SO}b), have a flattened spatial distribution along the major axis of the disk, and their \Vlsr\ gradients trace the disk rotation. We confirm the results of \citet{Bel14}, and the improvement in angular resolution (from $\approx$0\farcs4 to $\approx$0\farcs1) now allows the disk kinematics to be fully resolved. Figure~\ref{G035_PV} shows that all the molecular lines, and particularly the SO emission, present a Keplerian velocity profile. The archival data of \targ\ belong to the “Digging into the Interior of Hot Cores with ALMA” (DIHCA) project, a large program to study at high angular resolution the disk kinematics in \ $\approx$30 high-mass YSOs. \citet{Olg26} have recently completed the analysis of the gas kinematics for all the regions, and, specifically for \targ,  their fit of the Keplerian velocity profile of the CH$_3$CN emission yields a YSO mass of $\approx$20~\ms. This value is significantly higher than the previous estimate of 5--13~\ms\ by \citet{Bel14}, which was probably affected by the  insufficient angular resolution. 
 
Figure~\ref{G035_mol} shows that the emissions of the molecular lines, dust, and free-free continuum are basically overlapping in the disk. That is certainly due to insufficient angular resolution to resolve the disk structure vertically, since ionized and neutral gas cannot be cospatial. This is also indicated by the fact that the emission peaks of the neutral components, i.e., molecular lines and dust, are slightly displaced with respect to that of the free-free continuum (see Fig.\ref{G035_mol}). It is natural to assume that the molecular gas and dust trace the disk midplane, while higher levels of ionization are expected in the disk atmosphere owing to both the stellar irradiation of energetic (mainly far-ultraviolet, FUV) photons \citep{Gor15} and shocks from turbulent motions or the launch of a fast wind \citep{Pas23}. Regarding this point, we note that the radio emission of \targ\ is orders of magnitude below the level expected from ionization by the Lyman continuum of a zero-age-main-sequence (ZAMS) star of corresponding bolometric luminosity \citep{San18,San19b}, which favors shocks as the main route of gas ionization. However, the high temperatures of 300--500~K of the molecular gas derived from the fit of the CH$_3$OH and CH$_3$CN lines (see Table~\ref{mol_fit}) indicate that the disk is strongly heated by the stellar radiation and possibly even ionized at small radii. 

The SO and SO$_2$ emissions in \tari\ and \targ\ share a number of common properties. In both YSOs, the SO emission extends well above and below the disk and presents the clearest Keplerian-like rotation profile (see Figs.~\ref{I21_PV}a~and~\ref{G035_PV}a). As in \tari, also in \targ\ the PV plots of SO across the disk midplane and offset of the disk (see Fig.~\ref{G035_PV_off}) are fully consistent with each other, which ensures that SO is tracing the rotation profile of the gas in the wind. In both YSOs, SO$_2$ emerges from gas seen in projection close to the disk rotation axis that yields the steepest profile of velocity with position (see Figs.~\ref{I21_PV}d~and~\ref{G035_PV}d). Following the same reasonings as in Sect.~\ref{McwI}, we propose that the SO and SO$_2$ emissions trace a magneto-centrifugal wind in \targ, too. We provide further support to this hypothesis in Sect.~\ref{XvsD-G}.

\subsection{X-wind versus disk wind}
\label{XvsD}
So far, we have discussed the evidence for magneto-centrifugally launched winds in \tari\ and \targ, but we have not discriminated between the alternative scenarios of X-wind or DW. In the former case, the observed rotating molecular outflows should be ambient material entrained in a fast wide-angle X-wind launched at radii \ $\le 0.1$~au \citep{Lop19}. In the latter case, the gas would be ejected directly from an extended portion, $\sim$1--10~au, of the accretion disk.

\subsubsection{\tari}
\label{XvsD-I}
 
In \tari\ the emissions of CH$_3$CN and CH$_3$OH (see Fig.~\ref{I21_mol}, panels~a~and~b, respectively) wrap the radio jet inside a funnel-like pattern that is significantly wider than the longitudinal patterns of SO (see Fig.~\ref{SO}a), and HC$_3$N and SO$_2$ (see Fig.~\ref{I21_mol}, panels~c~and~d, respectively). In addition, the PV plots of CH$_3$CN and CH$_3$OH present elongated emission threading the second and fourth quadrants (see Fig.~\ref{I21_PV}), which is consistent with a rotating ring. The latter could actually be the kinematic structure responsible for the emission of the two molecules, since all the corresponding lines are optically thin (see Table~\ref{mol_fit}) and we would have seen the contribution of gas inside the rings otherwise.

Besides the diverse radial extension, there are also clear differences present in the structure of the molecular emissions. While the emissions of SO, HC$_3$N, and SO$_2$ are rather homogeneous, those of  CH$_3$CN and CH$_3$OH are irregular and concentrated in a few blobs with sizes of 500--1000~au (see Fig.~\ref{I21_mol}). All these molecular species are mainly excited in the YSO wind, as one can infer comparing the physical conditions derived at the positions of the YSO and outflow (see Table~\ref{mol_fit}). Except for CH$_3$CN, which certainly has a contribution from the denser and warmer disk gas toward the YSO, all the other molecules have comparable values for the excitation temperature and, in some cases, the column density, too, between the YSO and outflow positions. If the molecular emissions arise in shocks of the protostellar wind, the different structures can be explained by the varying nature of the shocks. The wider and irregular intensity distributions of CH$_3$CN and CH$_3$OH can be produced by a discrete number of relatively large (exceeding the beam size of 250~au) and strong, external shocks of the wind against dense clumps of the surrounding medium. The homogeneous emissions of SO, HC$_3$N, and SO$_2$ can result from a higher number of smaller ($\sim$10--100~au) weaker internal shocks. 

High abundances of CH$_3$OH are expected in the post-shock layers of C-type shocks for shock velocity  $V_{\rm s} \gtrsim$12~\kms\ sufficiently high for efficient sputtering of the CH$_3$OH molecules off the icy mantle of dust grains \citep{Burk19,Bai26}. For a pre-shock hydrogen number density, $n_{\rm H}$, of 10$^{5}$--10$^6$~cm$^{-3}$, C-type shocks yield post-shock depths of warm molecular gas of  $\approx$500--5000~au \citep[see, e.g.,][]{Pin93}, comparable with or larger than the irregular features observed in the CH$_3$CN and CH$_3$OH maps. Our excitation analysis confirms that the CH$_3$OH column density in the emission blobs far from the outflow axis is higher than along the outflow axis (see Table~\ref{mol_fit}), despite the total gas density is expected to diminish with radius. It is also notable that the CH$_3$CN column density keeps about the same close and far from the outflow axis, while that of SO and SO$_2$ decreases from small to large radii.

On the other hand, the recent calculations by \citet{vGel21}, accounting for both warm gas-phase chemistry and thermal desorption from dust grains, show that 10--100~au layers of enhanced SO and SO$_2$ abundance can be produced behind J-~and~CJ-type shocks propagating across an high-density, $n_{\rm H}$ $\sim$ 10$^{7}$~cm$^{-3}$, medium at a relatively low velocity, \ $V_{\rm s} \approx$ $1-10$~\kms. In particular, the enhancement of SO$_2$ is predicted to occur only for the upper range of shock velocities, \ $V_{\rm s} \approx$ $5-10$~\kms. In these shocks, while the gas reaches temperatures of a few thousand Kelvin at which more complex molecules (such as CH$_3$CN and CH$_3$OH) dissociate, the dust temperature, owing to efficient radiative cooling, stays lower than 70~K. Sublimation from the grain mantles of SO (at $\approx$37~K) and SO$_2$ (at $\approx$62~K) ices can occur, but not that of CH$_3$OH and H$_2$O ices requiring dust temperatures of 85--100~K \citep{Suu14}. 

The above comparison with shock models supports our interpretation of the spatial distribution and structure of the CH$_3$OH and SO and SO$_2$ emissions in \tari\ in terms of large external and small internal shocks of the protostellar wind, respectively. Then, only a fraction of the observed CH$_3$OH (and CH$_3$CN) can possibly correspond to ambient material entrained in the external shocks of the wind, but the inner gas traced by SO and SO$_2$ can stem directly from the disk, endorsing the case of a MHD DW. 

It is also interesting to note that the models by \citet{vGel21} require higher shock velocities for the formation of SO$_2$ with respect to SO. That agrees very well with the MHD DW interpretation, where the gas traced by SO$_2$, closer to the jet axis, is predicted to move faster than that emitting SO. Therefore, it is also natural to expect that the internal shocks of the flow producing SO$_2$ have higher velocities than those of SO. Following the same reasoning, at sufficiently large radii, the velocity of MHD DWs and corresponding internal shocks would become too low ($\le$ 1~\kms) for efficient production of SO, determining a maximum radius for the SO emission to be one of the preferential molecular tracers of DWs. Our analysis of the physical conditions supports both the models by \citet{vGel21} and the MHD DW interpretation, since the excitation temperature of SO$_2$ is found to be systematically higher than that of SO at all the explored positions (see Table~\ref{mol_fit}), in agreement with production in stronger shocks, and the continuous decrease in the SO$_2$ column density from near the YSO through small to large radii (see Table~\ref{mol_fit}) clearly indicates its higher concentration in faster gas closer to the outflow axis.

So far we have considered the case that the observed molecular emissions are mainly determined by processes in the local shocks of the wind, but, alternatively, they could mainly reflect the properties of the original gas of the primary wind. The radial distribution of the molecular emissions in \tari\ is reminiscent of what is found in low-mass YSOs, where, in an increasing number of cases, combined James Webb Space Telescope (JWST) and ALMA observations are now discovering fast collimated [Fe~\textsc{II}] jets lying inside a hollow cavity delimited by wider and slower H$_2$ emission, which is itself nested inside a still wider and slower CO wind \citep{Fed24,Tyc24,Pas25}. Such a nested morphology in velocity and chemistry of rotating winds at scales of a few hundred astronomical units is a distinctive characteristic of radially extended MHD DWs, in low-mass YSOs, at least, since it is very difficult to explain this as the result of gas entrainment \citep{Lop19,deV22}.

\subsubsection{\targ}
\label{XvsD-G}

In \targ\ the protostellar wind is traced by SO emission only. That already rules out the possibility that we are observing ambient gas entrained in a fast wide-angle X-wind as there is no reasonable explanation why all the other observed molecular emissions would not be present in the wind. However, before concluding in favor of a MHD DW, we need also to discuss the reason for the absence of these emissions if the wind material arises directly from the YSO disk.

As is pointed out in Sect.~\ref{Int}, the water maser 3D velocities appear to trace a rotating wind emerging from the disk. The high maser velocities, $\ge$20~\kms, favor a MHD DW rather than a photoevaporated DW, whose typical speeds are $\le$10~\kms\ \citep{Hol94,Ale14}.
Figure~\ref{SO}b (lower~plot) shows that most of the water masers are found near the peaks of the free-free continuum emission. Since water masers are typical shock tracers on scales of 1-10~au \citep{Hol13}, that reinforces the interpretation of the free-free continuum in terms of shock ionization. The radio continuum, comparable in size to the elongation of the water maser distribution, could trace the launch region of the protostellar wind, the water masers marking strong ionizing shocks of the wind against dense static parcels of ambient gas.

Magnetohydrodynamic disk winds are launched from the disk atmosphere \citep[][]{Pud07}, which could be found at significantly higher temperature than the disk midplane. The  temperature of the neutral gas of 300--500~K at radii of a few hundred astronomical units measured through the CH$_3$CN and CH$_3$OH lines (see Table~\ref{mol_fit}) is in good agreement with predictions of the most recent simulations of high-mass YSOs, which also consider radiation forces and photo-ionization feedback \citep[][see their Fig.~6]{Kui18}. If the disk atmosphere were heated by the energetic stellar photons of a $\approx$20~\ms\ star, its temperature would be 10$^3$--10$^4$~K, as indicated by models of photo-ionization of massive disks \citep{Hol94,Yor96}. In this specific case, the presence of an extended region of photo-ionized gas is unlikely, since the radio emission is orders of magnitude below the Lyman continuum level. The star could be in a bloating phase \citep{Hos09,Hos16}, predicted to occur within the range of YSO mass of 10--30~\ms\ \citep{Kui13,Kui18} and effectively observed for a $\approx$25~\ms\ YSO \citep{Pan25}, and its photospheric temperature could be too low to emit ionizing photons. 

In \targ, however, there is a strong observational evidence that the disk surface is shock-ionized. Therefore, it is plausible to assume that the disk atmosphere is sufficiently warm over the DW launch region for the molecular gas to be partially or fully dissociated. The dust, instead, owing to efficient radiative cooling, could still retain temperatures of a few tens of Kelvin. That could explain the lack of molecular emissions other than SO in the DW. We note that the SO gas-phase formation in J-type shocks requires only the presence of dust in the pre-shock gas, as all the relevant molecular reactants are directly obtained through desorption from the icy dust mantles \citep{vGel21}.

The SO$_2$ emission in \targ\ presents a compact arched structure contiguous to the disk (see Fig.~\ref{G035_mol}d).  That leads us to suppose that it could trace the base of the cavity of the DW and be excited in relatively strong shocks of the DW against the surrounding dense, likely infalling, gas. A similar interpretation holds also for the X-shape structure traced by the SO$_2$ emission that emerges from the rotating disk in G011.92$-$0.61 \citep[][see their Figs.~2~and~9]{Bay25}. 

In \targ, the lack of SO$_2$ in the DW at larger distances from the disk and close to the jet axis, as is instead observed in \tari, could still be linked with the paucity of the ultraviolet (UV) flux from the bloated star. 
In fact, the gas-phase formation of SO and SO$_2$ in J-shocks crucially depends on the strength of the local UV radiation field for the production of essential reactants via the photo-dissociation of \ H$_2$O, H$_2$S, and CH$_4$ \citep{vGel21}.
Models of massive star formation predict that the radiation from the star is preferentially beamed along the optically thin bipolar-outflow cavity \citep{Kru05,Kui10}, which in \tari\ could account for a sufficiently high UV-radiation flux close to the jet axis to form SO$_2$. On the other hand, if the source of ionization are exclusively shocks over the disk surface, one can expect that the diffuse UV radiation can affect only the DW gas closer to the disk, in agreement with the observed spatial distribution of the SO emission (see Fig.~\ref{SO}b, upper~plot), and is absorbed before reaching the high-velocity shocks along the jet axis.

\section{Conclusions}
\label{Con}
The POETS survey has investigated the launch region of protostellar outflows from luminous YSOs at scales of 10--100~au through VLBI observations of the water masers, finding good evidence for MHD DWs. Considering the fundamental role that MHD DWs could have in driving disk accretion and evolution, it is necessary to further test their presence and characterize their properties with complementary high-resolution millimeter interferometric observations. In our previous study of the POETS target G011.92$-$0.61 \citep{Bay25}, the MHD DW interpretation based on water maser 3D velocities has been fully confirmed via sensitive ALMA observations at a linear resolution of $\approx$100~au. In this article, we have focused on the YSOs \tari\ and \targ, two POETS targets of different mass, 5.6$\pm$2 and $\approx$20~\ms, respectively, associated with characteristic maser MHD DW configurations.
We have investigated the base of the protostellar outflows at scales of 100--1000~au in these two YSOs using the same molecular lines at comparable linear and velocity resolutions and sensitivity. 

In both YSOs, all the molecular emissions trace a clear \Vlsr\ gradient transversal to the disk rotation (and, in \tari, the radio jet) axis. In both sources, the SO emission is one of the most radially extended and presents the clearest Keplerian-like velocity  profile in the PV plots, while SO$_2$ is spatially compact and yields the steepest velocity-position profile. Our analysis of these kinematic findings supports the case of a wind magneto-centrifugally launched from the YSO disk. 

The wind in \tari\ harbors many molecules with a clear radial distribution, going from SO$_2$ and HC$_3$N, found close to the jet axis, through SO, covering larger radii, to CH$_3$CN and CH$_3$OH, with the maximum width. In \targ, only SO traces the wind, while the other molecular emissions are confined inside the disk. That already excludes the possibility of a X-wind, i.e., that the wind gas is entrained from the surrounding medium, as we should see other molecules, as well. Instead, if we are observing MHD DWs, i.e., if the wind gas is ejected directly from the disk, models of molecule production in shocks can explain both the radial distribution of molecular emissions in the protostellar wind of \tari\ and the absence of molecules, other than SO, in the \targ's wind. The latter is a consequence of the fact that the launch region of the wind in the disk atmosphere is shock-ionized and the molecular gas is likely partially or fully dissociated.

The results obtained in \tari\ and \targ, for two YSOs of significantly different mass, confirm the POETS conclusion that MHD DWs can be the common launching mechanism of protostellar winds in luminous YSOs. The gas kinematics traced by water masers at scales of 10--100~au and determined via thermal molecular lines at scales of 100--1000~au are very consistent. Considering also our recent findings in G011.92$-$0.61, we expect that SO and SO$_2$ emissions can be among the preferential tracers of MHD DWs in intermediate-~and~high-mass YSOs at small scales.

\begin{acknowledgements}
A.~P. acknowledges financial support from the UNAM-PAPIIT IN120226 grant, and the Sistema Nacional de Investigadores of SECIHTI, M\'exico.
A.~S-M.\ acknowledges support from the PID2023-146675NB grant funded by MCIN/AEI/10.13039/501100011033, and by the Spanish program Unidad de Excelencia María de Maeztu CEX2020-001058-M, financed by MCIN/AEI/10.13039/501100011033, and by the MaX-CSIC Excellence Award MaX4-SOMMA-ICE.
D.~S. was funded by the Deutsche Forschungsgemeinschaft (DFG, German Research Foundation) – project number: 550639632.
R.~E.~P. is supported by a Discovery Grant from the National Science and Engineering Research Council (NSERC) of Canada. 
R.~K. acknowledges financial support via the Heisenberg Research Grant funded by the Deutsche Forschungsgemeinschaft (DFG, German Research Foundation) under grant no.~KU 2849/9, project no.~445783058. 
\end{acknowledgements}

%

\begin{thebibliography}{67}
\expandafter\ifx\csname natexlab\endcsname\relax\def\natexlab#1{#1}\fi

\bibitem[{{Ahmadi} {et~al.}(2023){Ahmadi}, {Beuther}, {Bosco}, {Gieser},
  {Suri}, {Mottram}, {Kuiper}, {Henning}, {S{\'a}nchez-Monge}, {Linz},
  {Pudritz}, {Semenov}, {Winters}, {M{\"o}ller}, {Beltr{\'a}n}, {Csengeri},
  {Galv{\'a}n-Madrid}, {Johnston}, {Keto}, {Klaassen}, {Leurini}, {Longmore},
  {Lumsden}, {Maud}, {Moscadelli}, {Palau}, {Peters}, {Ragan}, {Urquhart},
  {Zhang}, \& {Zinnecker}}]{Ahm23}
{Ahmadi}, A., {Beuther}, H., {Bosco}, F., {et~al.} 2023, \aap, 677, A171

\bibitem[{{Alexander} {et~al.}(2014){Alexander}, {Pascucci}, {Andrews},
  {Armitage}, \& {Cieza}}]{Ale14}
{Alexander}, R., {Pascucci}, I., {Andrews}, S., {Armitage}, P., \& {Cieza}, L.
  2014, in Protostars and Planets VI, ed. H.~{Beuther}, R.~S. {Klessen}, C.~P.
  {Dullemond}, \& T.~{Henning}, 475

\bibitem[{{Baijot} {et~al.}(2026){Baijot}, {Groyne}, \& {De Becker}}]{Bai26}
{Baijot}, C., {Groyne}, M., \& {De Becker}, M. 2026, \aap, 705, A185

\bibitem[{{Bayandina} {et~al.}(2025){Bayandina}, {Moscadelli}, {Cesaroni},
  {Beltr{\'a}n}, {Sanna}, \& {Goddi}}]{Bay25}
{Bayandina}, O.~S., {Moscadelli}, L., {Cesaroni}, R., {et~al.} 2025, \aap, 694,
  A92

\bibitem[{{Beltr{\' a}n} {et~al.}(2004){Beltr{\' a}n}, {Cesaroni}, {Neri},
  {Codella}, {Furuya}, {Testi}, \& {Olmi}}]{Bel04}
{Beltr{\' a}n}, M.~T., {Cesaroni}, R., {Neri}, R., {et~al.} 2004, \apjl, 601,
  L187

\bibitem[{{Beltr{\'a}n} {et~al.}(2014){Beltr{\'a}n}, {S{\'a}nchez-Monge},
  {Cesaroni}, {Kumar}, {Galli}, {Walmsley}, {Etoka}, {Furuya}, {Moscadelli},
  {Stanke}, {van der Tak}, {Vig}, {Wang}, {Zinnecker}, {Elia}, \&
  {Schisano}}]{Bel14}
{Beltr{\'a}n}, M.~T., {S{\'a}nchez-Monge}, {\'A}., {Cesaroni}, R., {et~al.}
  2014, \aap, 571, A52

\bibitem[{{Beuther} {et~al.}(2018){Beuther}, {Mottram}, {Ahmadi}, {Bosco},
  {Linz}, {Henning}, {Klaassen}, {Winters}, {Maud}, {Kuiper}, {Semenov},
  {Gieser}, {Peters}, {Urquhart}, {Pudritz}, {Ragan}, {Feng}, {Keto},
  {Leurini}, {Cesaroni}, {Beltran}, {Palau}, {S{\'a}nchez-Monge},
  {Galvan-Madrid}, {Zhang}, {Schilke}, {Wyrowski}, {Johnston}, {Longmore},
  {Lumsden}, {Hoare}, {Menten}, \& {Csengeri}}]{Beu18}
{Beuther}, H., {Mottram}, J.~C., {Ahmadi}, A., {et~al.} 2018, \aap, 617, A100

\bibitem[{{Blandford} \& {Payne}(1982)}]{Bla82}
{Blandford}, R.~D. \& {Payne}, D.~G. 1982, \mnras, 199, 883

\bibitem[{{Briggs}(1995)}]{Bri95}
{Briggs}, D.~S. 1995, in Bulletin of the American Astronomical Society,
  Vol.~27, American Astronomical Society Meeting Abstracts, 1444

\bibitem[{{Burkhardt} {et~al.}(2019){Burkhardt}, {Shingledecker}, {Le Gal},
  {McGuire}, {Remijan}, \& {Herbst}}]{Burk19}
{Burkhardt}, A.~M., {Shingledecker}, C.~N., {Le Gal}, R., {et~al.} 2019, \apj,
  881, 32

\bibitem[{{CASA Team} {et~al.}(2022){CASA Team}, {Bean}, {Bhatnagar}, {Castro},
  {Donovan Meyer}, {Emonts}, {Garcia}, {Garwood}, {Golap}, {Gonzalez Villalba},
  {Harris}, {Hayashi}, {Hoskins}, {Hsieh}, {Jagannathan}, {Kawasaki},
  {Keimpema}, {Kettenis}, {Lopez}, {Marvil}, {Masters}, {McNichols},
  {Mehringer}, {Miel}, {Moellenbrock}, {Montesino}, {Nakazato}, {Ott}, {Petry},
  {Pokorny}, {Raba}, {Rau}, {Schiebel}, {Schweighart}, {Sekhar}, {Shimada},
  {Small}, {Steeb}, {Sugimoto}, {Suoranta}, {Tsutsumi}, {van Bemmel},
  {Verkouter}, {Wells}, {Xiong}, {Szomoru}, {Griffith}, {Glendenning}, \&
  {Kern}}]{CASA22}
{CASA Team}, {Bean}, B., {Bhatnagar}, S., {et~al.} 2022, \pasp, 134, 114501

\bibitem[{{Cesaroni} {et~al.}(1999){Cesaroni}, {Felli}, {Jenness}, {Neri},
  {Olmi}, {Robberto}, {Testi}, \& {Walmsley}}]{Ces99}
{Cesaroni}, R., {Felli}, M., {Jenness}, T., {et~al.} 1999, \aap, 345, 949

\bibitem[{{De Simone} {et~al.}(2024){De Simone}, {Podio}, {Chahine}, {Codella},
  {Chandler}, {Ceccarelli}, {L{\'o}pez-Sepulcre}, {Loinard}, {Svoboda},
  {Sakai}, {Johnstone}, {M{\'e}nard}, {Aikawa}, {Bouvier}, {Sabatini},
  {Miotello}, {Vastel}, {Cuello}, {Bianchi}, {Caselli}, {Caux}, {Hanawa},
  {Herbst}, {Segura-Cox}, {Zhang}, \& {Yamamoto}}]{DeS24}
{De Simone}, M., {Podio}, L., {Chahine}, L., {et~al.} 2024, \aap, 686, L13

\bibitem[{{de Valon} {et~al.}(2022){de Valon}, {Dougados}, {Cabrit}, {Louvet},
  {Zapata}, \& {Mardones}}]{deV22}
{de Valon}, A., {Dougados}, C., {Cabrit}, S., {et~al.} 2022, \aap, 668, A78

\bibitem[{{Endres} {et~al.}(2016){Endres}, {Schlemmer}, {Schilke}, {Stutzki},
  \& {M{\"u}ller}}]{End16}
{Endres}, C.~P., {Schlemmer}, S., {Schilke}, P., {Stutzki}, J., \&
  {M{\"u}ller}, H. S.~P. 2016, Journal of Molecular Spectroscopy, 327, 95

\bibitem[{{Federman} {et~al.}(2024){Federman}, {Megeath}, {Rubinstein},
  {Gutermuth}, {Narang}, {Tyagi}, {Manoj}, {Anglada}, {Atnagulov}, {Beuther},
  {Bourke}, {Brunken}, {Caratti o Garatti}, {Evans}, {Fischer}, {Furlan},
  {Green}, {Habel}, {Hartmann}, {Karnath}, {Klaassen}, {Linz}, {Looney},
  {Osorio}, {Muzerolle Page}, {Nazari}, {Pokhrel}, {Rahatgaonkar}, {Rocha},
  {Sheehan}, {Slavicinska}, {Stanke}, {Stutz}, {Tobin}, {Tychoniec}, {Van
  Dishoeck}, {Watson}, {Wolk}, \& {Yang}}]{Fed24}
{Federman}, S.~A., {Megeath}, S.~T., {Rubinstein}, A.~E., {et~al.} 2024, \apj,
  966, 41

\bibitem[{{Gieser} {et~al.}(2021){Gieser}, {Beuther}, {Semenov}, {Ahmadi},
  {Suri}, {M{\"o}ller}, {Beltr{\'a}n}, {Klaassen}, {Zhang}, {Urquhart},
  {Henning}, {Feng}, {Galv{\'a}n-Madrid}, {de Souza Magalh{\~a}es},
  {Moscadelli}, {Longmore}, {Leurini}, {Kuiper}, {Peters}, {Menten},
  {Csengeri}, {Fuller}, {Wyrowski}, {Lumsden}, {S{\'a}nchez-Monge}, {Maud},
  {Linz}, {Palau}, {Schilke}, {Pety}, {Pudritz}, {Winters}, \&
  {Pi{\'e}tu}}]{Gie21}
{Gieser}, C., {Beuther}, H., {Semenov}, D., {et~al.} 2021, \aap, 648, A66

\bibitem[{{Gorti} {et~al.}(2015){Gorti}, {Hollenbach}, \& {Dullemond}}]{Gor15}
{Gorti}, U., {Hollenbach}, D., \& {Dullemond}, C.~P. 2015, \apj, 804, 29

\bibitem[{{H{\"o}gbom}(1974)}]{Hog74}
{H{\"o}gbom}, J.~A. 1974, \aaps, 15, 417

\bibitem[{{Hollenbach} {et~al.}(2013){Hollenbach}, {Elitzur}, \&
  {McKee}}]{Hol13}
{Hollenbach}, D., {Elitzur}, M., \& {McKee}, C.~F. 2013, \apj, 773, 70

\bibitem[{{Hollenbach} {et~al.}(1994){Hollenbach}, {Johnstone}, {Lizano}, \&
  {Shu}}]{Hol94}
{Hollenbach}, D., {Johnstone}, D., {Lizano}, S., \& {Shu}, F. 1994, \apj, 428,
  654

\bibitem[{{Hosokawa} {et~al.}(2016){Hosokawa}, {Hirano}, {Kuiper}, {Yorke},
  {Omukai}, \& {Yoshida}}]{Hos16}
{Hosokawa}, T., {Hirano}, S., {Kuiper}, R., {et~al.} 2016, \apj, 824, 119

\bibitem[{{Hosokawa} \& {Omukai}(2009)}]{Hos09}
{Hosokawa}, T. \& {Omukai}, K. 2009, \apj, 691, 823

\bibitem[{{Kadam} {et~al.}(2025){Kadam}, {Vorobyov}, {Woitke}, {Basu}, \& {van
  Terwisga}}]{Kad25}
{Kadam}, K., {Vorobyov}, E., {Woitke}, P., {Basu}, S., \& {van Terwisga}, S.
  2025, \aap, 695, A167

\bibitem[{{K\"{o}nigl} \& {Pudritz}(2000)}]{Kon00}
{K\"{o}nigl}, A. \& {Pudritz}, R.~E. 2000, Protostars and Planets IV, 759

\bibitem[{{Krumholz} {et~al.}(2005){Krumholz}, {McKee}, \& {Klein}}]{Kru05}
{Krumholz}, M.~R., {McKee}, C.~F., \& {Klein}, R.~I. 2005, \apjl, 618, L33

\bibitem[{{Kuiper} \& {Hosokawa}(2018)}]{Kui18}
{Kuiper}, R. \& {Hosokawa}, T. 2018, \aap, 616, A101

\bibitem[{{Kuiper} {et~al.}(2010){Kuiper}, {Klahr}, {Beuther}, \&
  {Henning}}]{Kui10}
{Kuiper}, R., {Klahr}, H., {Beuther}, H., \& {Henning}, T. 2010, \apj, 722,
  1556

\bibitem[{{Kuiper} \& {Yorke}(2013)}]{Kui13}
{Kuiper}, R. \& {Yorke}, H.~W. 2013, \apj, 763, 104

\bibitem[{{Levenberg}(1944)}]{Lev44}
{Levenberg}, K. 1944, Quarterly of Applied Mathematics, 2, 64

\bibitem[{{L{\'o}pez-V{\'a}zquez} {et~al.}(2019){L{\'o}pez-V{\'a}zquez},
  {Cant{\'o}}, \& {Lizano}}]{Lop19}
{L{\'o}pez-V{\'a}zquez}, J.~A., {Cant{\'o}}, J., \& {Lizano}, S. 2019, \apj,
  879, 42

\bibitem[{{Lynden-Bell} \& {Pringle}(1974)}]{Lyn74}
{Lynden-Bell}, D. \& {Pringle}, J.~E. 1974, \mnras, 168, 603

\bibitem[{{Marquardt}(1963)}]{Mar63}
{Marquardt}, D. 1963, Journal of the Society for Industrial and Applied
  Mathematics, 11, 431

\bibitem[{{Mart{\'\i}n} {et~al.}(2019){Mart{\'\i}n}, {Mart{\'\i}n-Pintado},
  {Blanco-S{\'a}nchez}, {Rivilla}, {Rodr{\'\i}guez-Franco}, \&
  {Rico-Villas}}]{Mar19}
{Mart{\'\i}n}, S., {Mart{\'\i}n-Pintado}, J., {Blanco-S{\'a}nchez}, C.,
  {et~al.} 2019, \aap, 631, A159

\bibitem[{{Moscadelli} {et~al.}(2021){Moscadelli}, {Beuther}, {Ahmadi},
  {Gieser}, {Massi}, {Cesaroni}, {S{\'a}nchez-Monge}, {Bacciotti},
  {Beltr{\'a}n}, {Csengeri}, {Galv{\'a}n-Madrid}, {Henning}, {Klaassen},
  {Kuiper}, {Leurini}, {Longmore}, {Maud}, {M{\"o}ller}, {Palau}, {Peters},
  {Pudritz}, {Sanna}, {Semenov}, {Urquhart}, {Winters}, \& {Zinnecker}}]{Mos21}
{Moscadelli}, L., {Beuther}, H., {Ahmadi}, A., {et~al.} 2021, \aap, 647, A114

\bibitem[{{Moscadelli} {et~al.}(2011){Moscadelli}, {Cesaroni}, {Rioja},
  {Dodson}, \& {Reid}}]{Mos11a}
{Moscadelli}, L., {Cesaroni}, R., {Rioja}, M.~J., {Dodson}, R., \& {Reid},
  M.~J. 2011, \aap, 526, A66+

\bibitem[{{Moscadelli} {et~al.}(2007){Moscadelli}, {Goddi}, {Cesaroni},
  {Beltr{\'a}n}, \& {Furuya}}]{Mos07}
{Moscadelli}, L., {Goddi}, C., {Cesaroni}, R., {Beltr{\'a}n}, M.~T., \&
  {Furuya}, R.~S. 2007, \aap, 472, 867

\bibitem[{{Moscadelli} {et~al.}(2024){Moscadelli}, {Oliva}, {Sanna}, {Surcis},
  \& {Bayandina}}]{Mos24}
{Moscadelli}, L., {Oliva}, A., {Sanna}, A., {Surcis}, G., \& {Bayandina}, O.
  2024, \aap, 690, A81

\bibitem[{{Moscadelli} {et~al.}(2016){Moscadelli}, {S{\'a}nchez-Monge},
  {Goddi}, {Li}, {Sanna}, {Cesaroni}, {Pestalozzi}, {Molinari}, \&
  {Reid}}]{Mos16}
{Moscadelli}, L., {S{\'a}nchez-Monge}, {\'A}., {Goddi}, C., {et~al.} 2016,
  \aap, 585, A71

\bibitem[{{Moscadelli} {et~al.}(2022){Moscadelli}, {Sanna}, {Beuther}, {Oliva},
  \& {Kuiper}}]{Mos22}
{Moscadelli}, L., {Sanna}, A., {Beuther}, H., {Oliva}, A., \& {Kuiper}, R.
  2022, Nature Astronomy, 6, 1068

\bibitem[{{Moscadelli} {et~al.}(2019{\natexlab{a}}){Moscadelli}, {Sanna},
  {Cesaroni}, {Rivilla}, {Goddi}, \& {Rygl}}]{Mos19}
{Moscadelli}, L., {Sanna}, A., {Cesaroni}, R., {et~al.} 2019{\natexlab{a}},
  \aap, 622, A206

\bibitem[{{Moscadelli} {et~al.}(2019{\natexlab{b}}){Moscadelli}, {Sanna},
  {Goddi}, {Krishnan}, {Massi}, \& {Bacciotti}}]{Mos19b}
{Moscadelli}, L., {Sanna}, A., {Goddi}, C., {et~al.} 2019{\natexlab{b}}, \aap,
  631, A74

\bibitem[{{M{\"u}ller} {et~al.}(2001){M{\"u}ller}, {Thorwirth}, {Roth}, \&
  {Winnewisser}}]{Mue01}
{M{\"u}ller}, H.~S.~P., {Thorwirth}, S., {Roth}, D.~A., \& {Winnewisser}, G.
  2001, \aap, 370, L49

\bibitem[{{Narang} {et~al.}(2026){Narang}, {Tyagi}, {Ohashi}, {Manoj},
  {Megeath}, {Tobin}, {Van Dishoeck}, {Evans}, {Watson}, {Caratti o Garatti},
  {J{\o}rgensen}, {Gutermuth}, {Aso}, {Beuther}, {Looney}, {Neufeld},
  {Anglada}, {Osorio}, {Rubinstein}, {Federman}, {Hartmann}, {Nazari},
  {Karnath}, {Linz}, {Stanke}, {Bourke}, {Yang}, {Kuiper}, {Green}, {Klaassen},
  {Zakri}, {Habel}, {Brunken}, {Muzerolle}, {Slavicinska}, {Stutz},
  {Tychoniec}, {Wolk}, {Rocha}, \& {Fischer}}]{Nar26}
{Narang}, M., {Tyagi}, H., {Ohashi}, N., {et~al.} 2026, \apj, 1000, 184

\bibitem[{{Olguin} {et~al.}(2026){Olguin}, {Sanhueza}, {Oya}, {Ginsburg},
  {Beltr{\'a}n}, {Morii}, {Galv{\'a}n-Madrid}, {Chen}, {Luo}, {Tanaka},
  {Zhang}, {Cheng}, {Nakamura}, {Li}, {Taniguchi}, {Garay}, {Zhang}, {Saito},
  {Sakai}, {Lu}, {Weng}, \& {Guzm{\'a}n}}]{Olg26}
{Olguin}, F.~A., {Sanhueza}, P., {Oya}, Y., {et~al.} 2026, arXiv e-prints,
  arXiv:2601.15371

\bibitem[{{Pandey} {et~al.}(2025){Pandey}, {Palau}, {Serna}, {Kuiper},
  {S{\'a}nchez-Monge}, {Sharma}, {Sahai}, {Contreras}, {Hern{\'a}ndez},
  {Rom{\'a}n-Z{\'u}{\~n}iga}, \& {Rodler}}]{Pan25}
{Pandey}, R., {Palau}, A., {Serna}, J., {et~al.} 2025, \mnras, 541, 3772

\bibitem[{{Pascucci} {et~al.}(2025){Pascucci}, {Beck}, {Cabrit}, {Bajaj},
  {Edwards}, {Louvet}, {Najita}, {Skinner}, {Gorti}, {Salyk}, {Brittain},
  {Krijt}, {Muzerolle Page}, {Ruaud}, {Schwarz}, {Semenov}, {Duch{\^e}ne}, \&
  {Villenave}}]{Pas25}
{Pascucci}, I., {Beck}, T.~L., {Cabrit}, S., {et~al.} 2025, Nature Astronomy,
  9, 81

\bibitem[{{Pascucci} {et~al.}(2023){Pascucci}, {Cabrit}, {Edwards}, {Gorti},
  {Gressel}, \& {Suzuki}}]{Pas23}
{Pascucci}, I., {Cabrit}, S., {Edwards}, S., {et~al.} 2023, in Astronomical
  Society of the Pacific Conference Series, Vol. 534, Protostars and Planets
  VII, ed. S.~{Inutsuka}, Y.~{Aikawa}, T.~{Muto}, K.~{Tomida}, \& M.~{Tamura},
  567

\bibitem[{{Pelletier} \& {Pudritz}(1992)}]{Pel92}
{Pelletier}, G. \& {Pudritz}, R.~E. 1992, \apj, 394, 117

\bibitem[{{Pineau des Forets} {et~al.}(1993){Pineau des Forets}, {Roueff},
  {Schilke}, \& {Flower}}]{Pin93}
{Pineau des Forets}, G., {Roueff}, E., {Schilke}, P., \& {Flower}, D.~R. 1993,
  \mnras, 262, 915

\bibitem[{{Pudritz} {et~al.}(2025){Pudritz}, {Cridland}, {Inglis}, \&
  {Alessi}}]{Pud25}
{Pudritz}, R.~E., {Cridland}, A.~J., {Inglis}, J., \& {Alessi}, M. 2025, arXiv
  e-prints, arXiv:2505.22724

\bibitem[{{Pudritz} {et~al.}(2007){Pudritz}, {Ouyed}, {Fendt}, \&
  {Brandenburg}}]{Pud07}
{Pudritz}, R.~E., {Ouyed}, R., {Fendt}, C., \& {Brandenburg}, A. 2007, in
  Protostars and Planets V, ed. B.~{Reipurth}, D.~{Jewitt}, \& K.~{Keil}, 277

\bibitem[{{S{\'a}nchez-Monge} {et~al.}(2018){S{\'a}nchez-Monge}, {Schilke},
  {Ginsburg}, {Cesaroni}, \& {Schmiedeke}}]{Sanc18}
{S{\'a}nchez-Monge}, {\'A}., {Schilke}, P., {Ginsburg}, A., {Cesaroni}, R., \&
  {Schmiedeke}, A. 2018, \aap, 609, A101

\bibitem[{{Sanna} {et~al.}(2019{\natexlab{a}}){Sanna}, {K{\"o}lligan},
  {Moscadelli}, {Kuiper}, {Cesaroni}, {Pillai}, {Menten}, {Zhang}, {Caratti o
  Garatti}, {Goddi}, {Leurini}, \& {Carrasco-Gonz{\'a}lez}}]{San19a}
{Sanna}, A., {K{\"o}lligan}, A., {Moscadelli}, L., {et~al.} 2019{\natexlab{a}},
  \aap, 623, A77

\bibitem[{{Sanna} {et~al.}(2010){Sanna}, {Moscadelli}, {Cesaroni}, {Tarchi},
  {Furuya}, \& {Goddi}}]{San10b}
{Sanna}, A., {Moscadelli}, L., {Cesaroni}, R., {et~al.} 2010, \aap, 517, A78+

\bibitem[{{Sanna} {et~al.}(2019{\natexlab{b}}){Sanna}, {Moscadelli}, {Goddi},
  {Beltr{\'a}n}, {Brogan}, {Caratti o Garatti}, {Carrasco-Gonz{\'a}lez},
  {Hunter}, {Massi}, \& {Padovani}}]{San19b}
{Sanna}, A., {Moscadelli}, L., {Goddi}, C., {et~al.} 2019{\natexlab{b}}, \aap,
  623, L3

\bibitem[{{Sanna} {et~al.}(2018){Sanna}, {Moscadelli}, {Goddi}, {Krishnan}, \&
  {Massi}}]{San18}
{Sanna}, A., {Moscadelli}, L., {Goddi}, C., {Krishnan}, V., \& {Massi}, F.
  2018, \aap, 619, A107

\bibitem[{{Shakura} \& {Sunyaev}(1973)}]{Sha73}
{Shakura}, N.~I. \& {Sunyaev}, R.~A. 1973, \aap, 24, 337

\bibitem[{{Shu} {et~al.}(1994){Shu}, {Najita}, {Ostriker}, {Wilkin}, {Ruden},
  \& {Lizano}}]{Shu94}
{Shu}, F., {Najita}, J., {Ostriker}, E., {et~al.} 1994, \apj, 429, 781

\bibitem[{{Suutarinen} {et~al.}(2014){Suutarinen}, {Kristensen}, {Mottram},
  {Fraser}, \& {van Dishoeck}}]{Suu14}
{Suutarinen}, A.~N., {Kristensen}, L.~E., {Mottram}, J.~C., {Fraser}, H.~J., \&
  {van Dishoeck}, E.~F. 2014, \mnras, 440, 1844

\bibitem[{{Tabone} {et~al.}(2017){Tabone}, {Cabrit}, {Bianchi}, {Ferreira},
  {Pineau des For{\^e}ts}, {Codella}, {Gusdorf}, {Gueth}, {Podio}, \&
  {Chapillon}}]{Tab17}
{Tabone}, B., {Cabrit}, S., {Bianchi}, E., {et~al.} 2017, \aap, 607, L6

\bibitem[{{Tabone} {et~al.}(2020){Tabone}, {Cabrit}, {Pineau des For{\^e}ts},
  {Ferreira}, {Gusdorf}, {Podio}, {Bianchi}, {Chapillon}, {Codella}, \&
  {Gueth}}]{Tab20}
{Tabone}, B., {Cabrit}, S., {Pineau des For{\^e}ts}, G., {et~al.} 2020, \aap,
  640, A82

\bibitem[{{Tychoniec} {et~al.}(2024){Tychoniec}, {van Gelder}, {van Dishoeck},
  {Francis}, {Rocha}, {Caratti o Garatti}, {Beuther}, {Gieser}, {Justtanont},
  {Linnartz}, {Le Gouellec}, {Perotti}, {Devaraj}, {Tabone}, {Ray}, {Brunken},
  {Chen}, {Kavanagh}, {Klaassen}, {Slavicinska}, {G{\"u}del}, \&
  {{\"O}stlin}}]{Tyc24}
{Tychoniec}, {\L}., {van Gelder}, M.~L., {van Dishoeck}, E.~F., {et~al.} 2024,
  \aap, 687, A36

\bibitem[{{van Gelder} {et~al.}(2021){van Gelder}, {Tabone}, {van Dishoeck}, \&
  {Godard}}]{vGel21}
{van Gelder}, M.~L., {Tabone}, B., {van Dishoeck}, E.~F., \& {Godard}, B. 2021,
  \aap, 653, A159

\bibitem[{{Wu} {et~al.}(2014){Wu}, {Sato}, {Reid}, {Moscadelli}, {Zhang}, {Xu},
  {Brunthaler}, {Menten}, {Dame}, \& {Zheng}}]{Wu14}
{Wu}, Y.~W., {Sato}, M., {Reid}, M.~J., {et~al.} 2014, \aap, 566, A17

\bibitem[{{Xu} {et~al.}(2013){Xu}, {Li}, {Reid}, {Menten}, {Zheng},
  {Brunthaler}, {Moscadelli}, {Dame}, \& {Zhang}}]{Xu13}
{Xu}, Y., {Li}, J.~J., {Reid}, M.~J., {et~al.} 2013, \apj, 769, 15

\bibitem[{{Yorke} \& {Welz}(1996)}]{Yor96}
{Yorke}, H.~W. \& {Welz}, A. 1996, \aap, 315, 555

\end{thebibliography}


\begin{appendix}




\clearpage
\onecolumn
\section{Fit of the molecular emission}
\label{Fit_ap}

\begin{figure}[ht!]
  \centering
    \includegraphics[width=0.5\hsize]{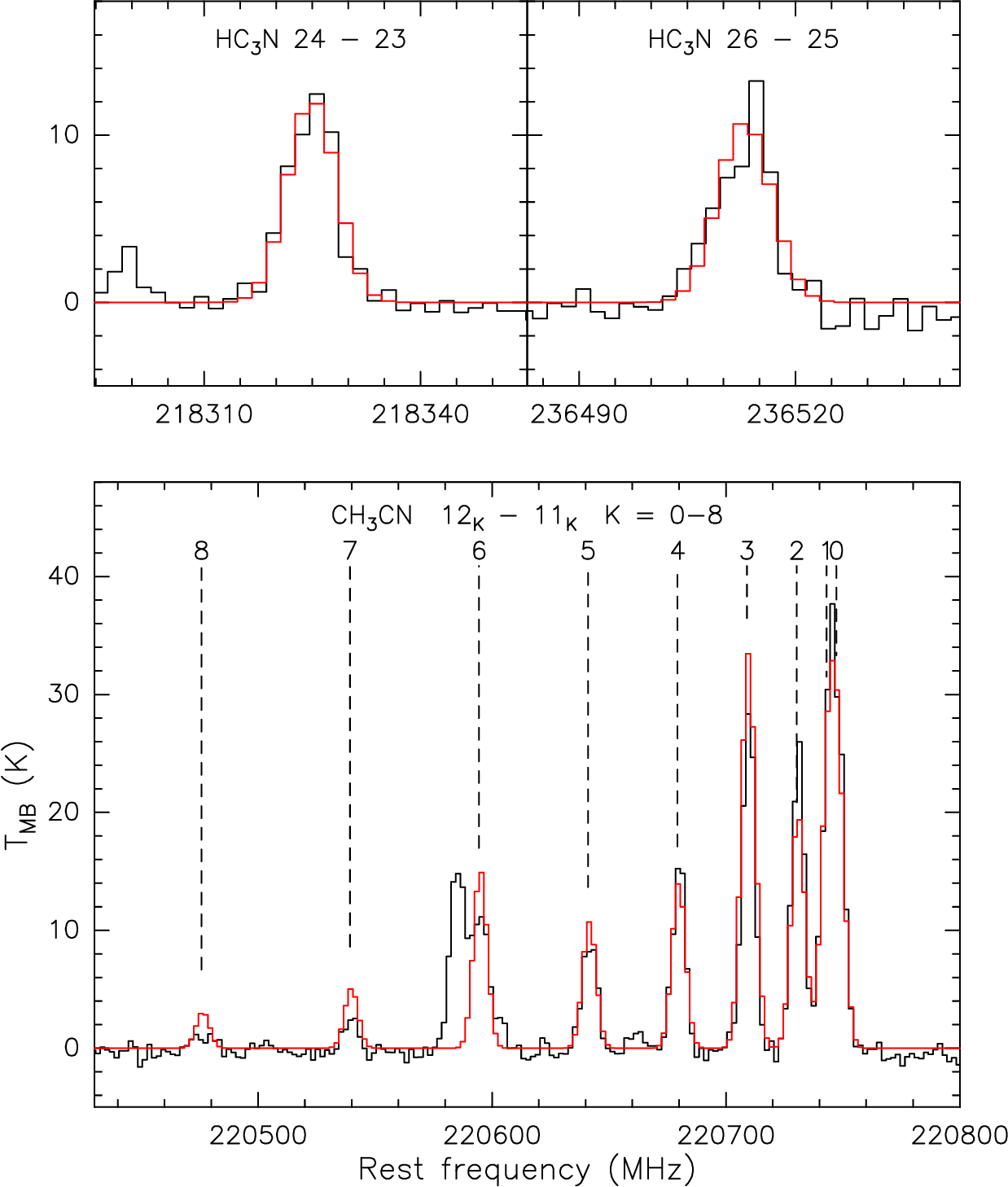} 
    \caption{Fit of the molecular emission in \tari. Each panel shows the observed (black) and fit (red) spectrum for a single or multiple molecular transitions, as labeled. Panels that refer to the same molecular species are grouped together.}
    \label{I21_fit_ch3cn-hc3n}
   \end{figure}

\begin{figure*}
  \centering
    \includegraphics[width=0.9\textwidth]{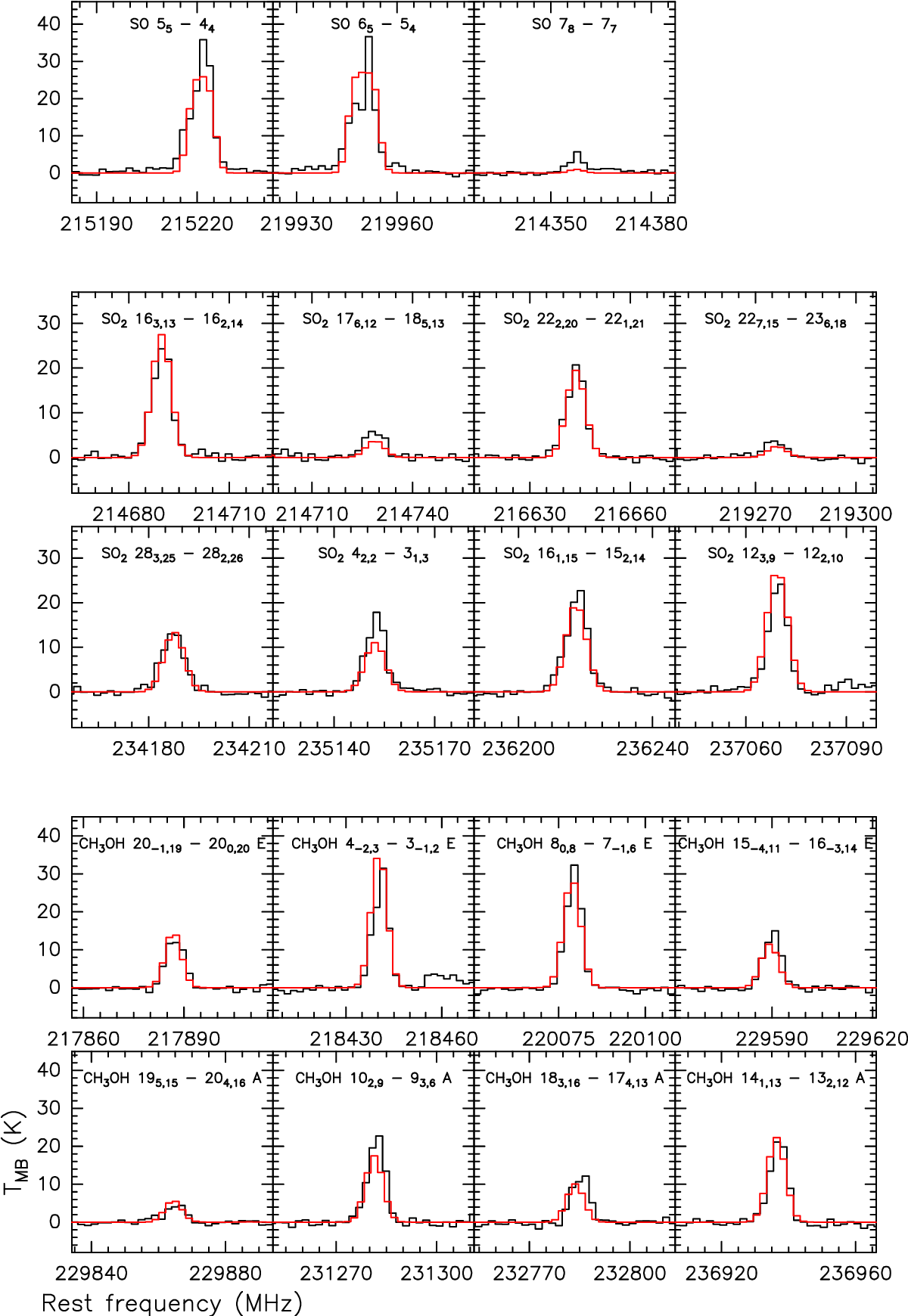} 
    \caption{Same as for Fig.~\ref{I21_fit_ch3cn-hc3n}.}
    \label{I21_fit}
   \end{figure*}

\begin{figure*}
  \centering
    \includegraphics[width=0.9\textwidth]{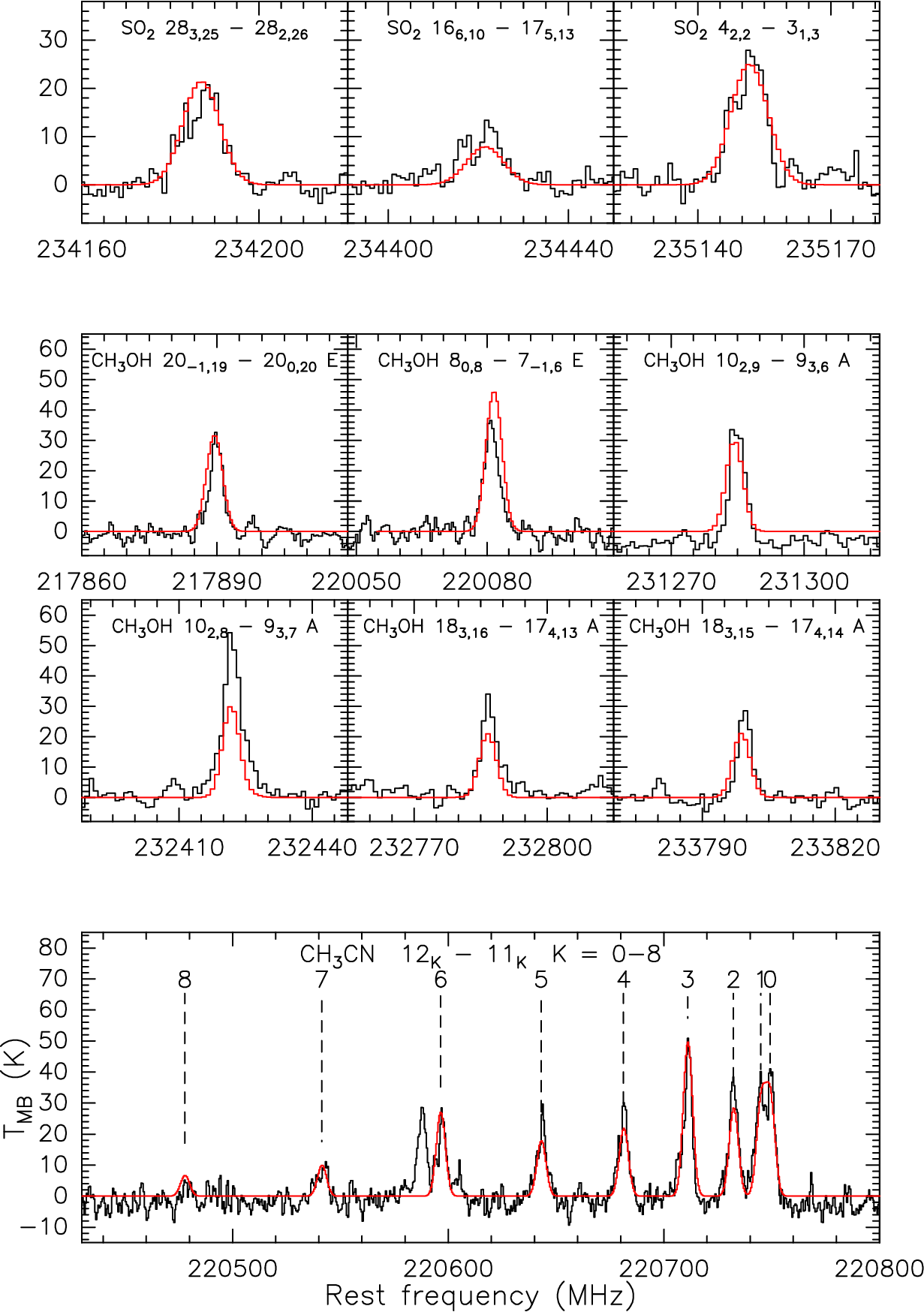} 
    \caption{Fit of the molecular emission in \targ. Each panel shows the observed (black) and fit (red) spectrum for a single or multiple molecular transitions, as labeled. Panels that refer to the same molecular species are grouped together.}
    \label{G035_fit}
   \end{figure*}

For each source, position and molecule reported in Table~\ref{mol_fit}, we have identified and fit with \textsc{MADCUBA} all the unblended transitions within the observed frequency range. The result of the fit for a subset of the more unblended and intense lines of each molecular species is presented in Figs.~\ref{I21_fit_ch3cn-hc3n}, \ref{I21_fit}, and~\ref{G035_fit}.  

\FloatBarrier

\onecolumn

\section{PV plots of the SO emission along the disk rotation axis}
\label{PV_ap}

\begin{figure}[ht!]
  \sidecaption
   \includegraphics[width=0.41\hsize]{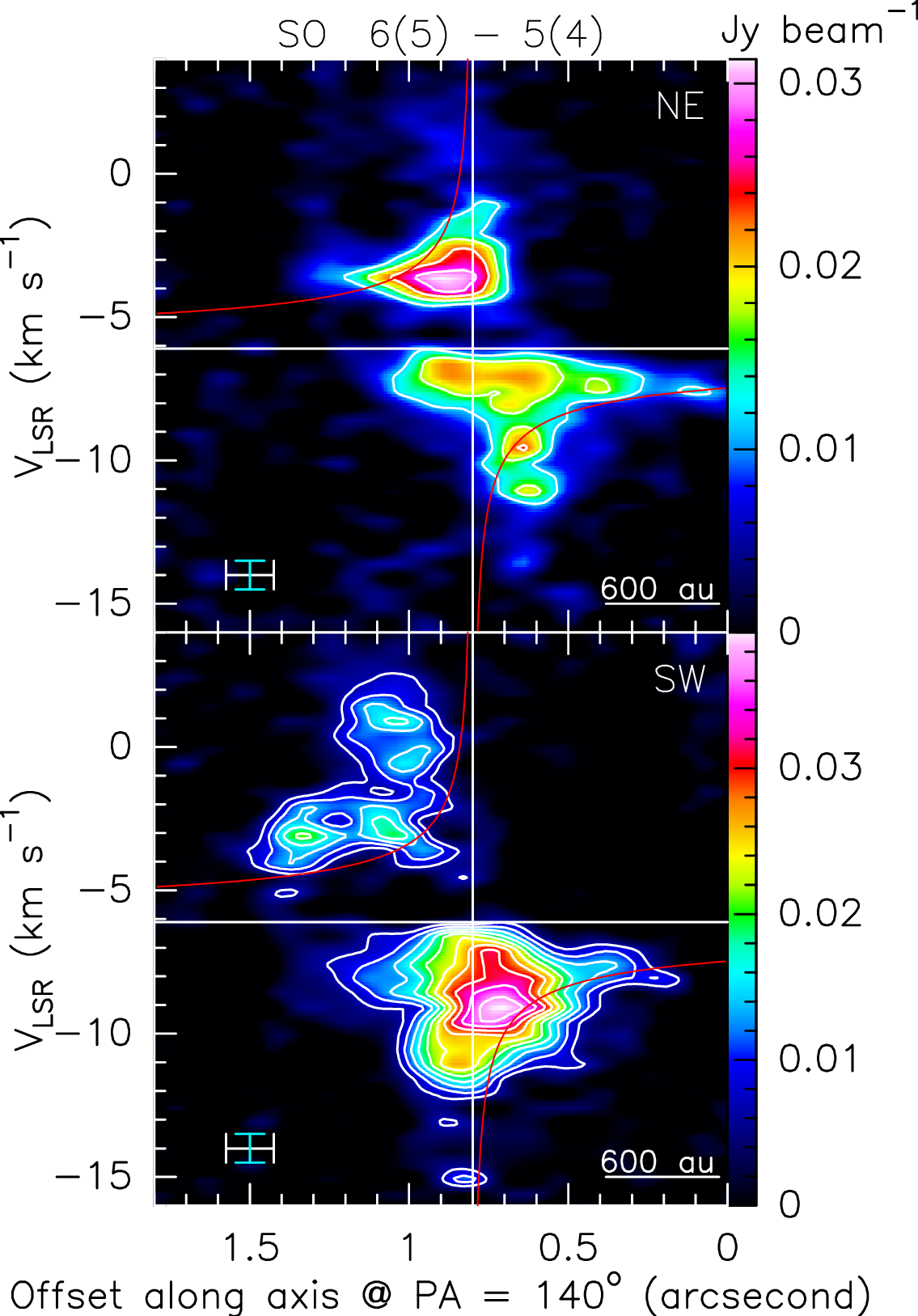} 
    \caption{PV plots of the SO emission in \tari\ at positions offset from the disk midplane. The upper and lower panels present the PV plots of the SO 6$_5$--5$_4$ line along the direction at PA = 140\degree \ (marked by the dashed black line in Figs.~\ref{SO}~and~\ref{I21_mol}) at positions along the disk rotation axis 0\farcs2 (i.e., the beam major axis) above (northeast) and 0\farcs2 below (southwest) the disk midplane, respectively. PV plots are shown with color maps and white contours, with contour levels ranging from $\left| I_{\rm min} \right|/2$ \ to \ $I_{\rm max}$ \ at steps of \ $\left| I_{\rm min} \right|/4$, where \ $I_{\rm min}$ \ and \  $I_{\rm max}$ \ are the minimum and maximum of the map, respectively. 
In each panel, the vertical and horizontal white axes denote the positional offset ($\approx$ 0\farcs8) of the YSO relative to the phase center of the observations and YSO \Vlsr\ ($\approx$ $-$6.1~\kms), respectively.   
For comparison, the red curve shows the Keplerian velocity profile for a central mass of 6~\ms. In the lower left corner of the panels, the vertical cyan and  horizontal white error bars indicate the velocity and spatial resolutions, respectively.}
    \label{I21_PV_off}
   \end{figure}

\begin{figure}[ht!]
\sidecaption
    \includegraphics[width=0.41\hsize]{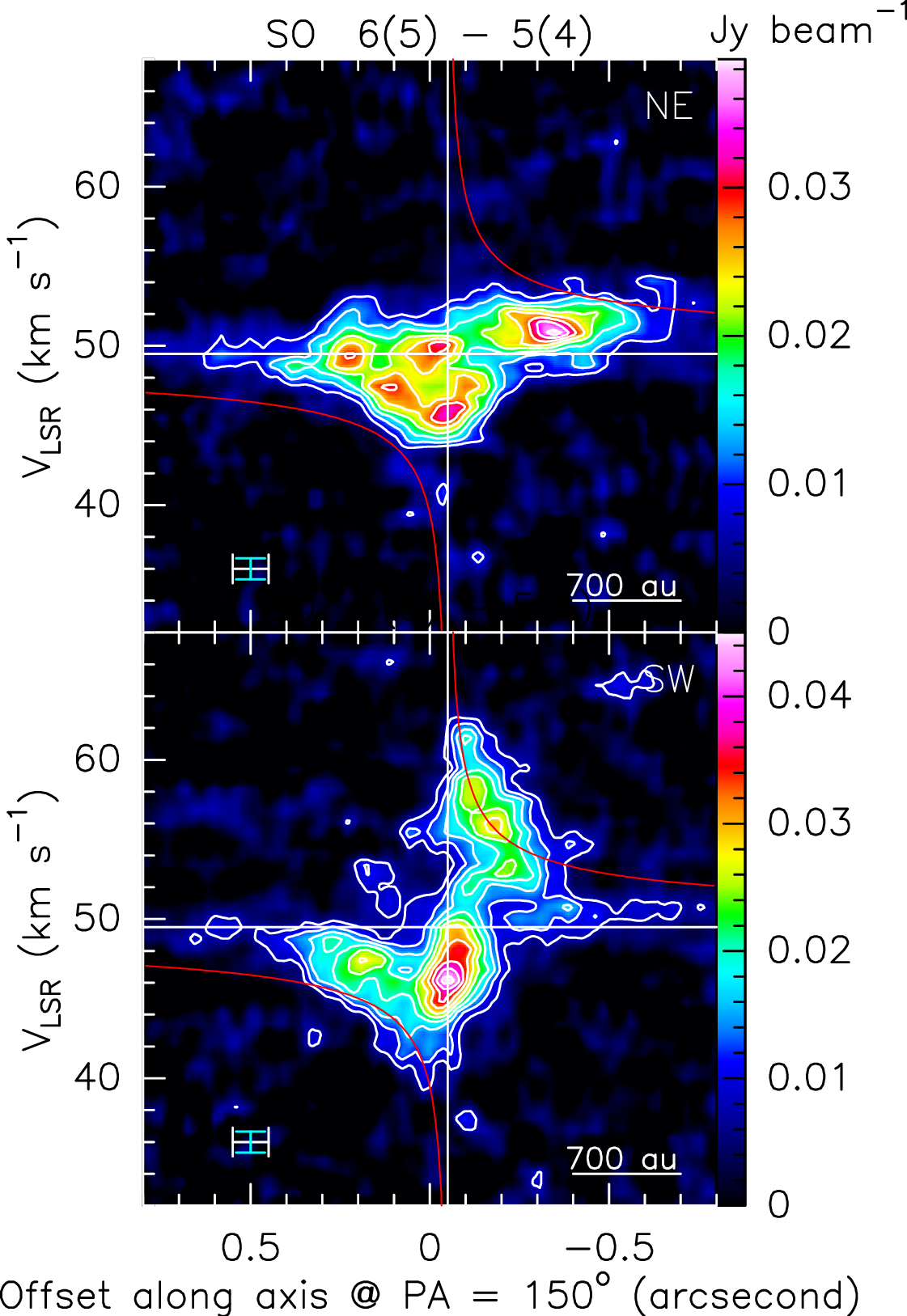} 
    \caption{PV plots of the SO emission in \targ\ at positions offset from the disk midplane. The upper and lower panels present the PV plots of the SO 6$_5$--5$_4$ line along the direction at PA = 150\degree \ (marked by the dashed black line in Figs.~\ref{SO}~and~\ref{G035_mol}) at positions along the disk rotation axis 0\farcs12 (i.e., the beam major axis) above (northeast) and 0\farcs12 below (southwest) the disk midplane, respectively. PV plots are shown with color maps and white contours, with contour levels ranging from $\left| I_{\rm min} \right|/2$ \ to \ $I_{\rm max}$ \ at steps of \ $\left| I_{\rm min} \right|/4$, where \ $I_{\rm min}$ \ and \  $I_{\rm max}$ \ are the minimum and maximum of the map, respectively. 
In each panel, the vertical and horizontal white axes denote the positional offset ($\approx$ $-$0\farcs05) of the YSO relative to the phase center of the observations and YSO \Vlsr\ ($\approx$ 49.5~\kms), respectively.   
For comparison, the red curve shows the Keplerian velocity profile for a central mass of 20~\ms. In the lower left corner of the panels, the vertical cyan and  horizontal white error bars indicate the velocity and spatial resolutions, respectively.}
    \label{G035_PV_off}
   \end{figure}

Figures~\ref{I21_PV_off}~and~\ref{G035_PV_off} report the PV plots of the SO 6$_5$--5$_4$ line at positions along the disk rotation axis above (northeast) and below (southwest) the disk midplane for \tari\ and \targ, respectively.

\FloatBarrier

\twocolumn

\end{appendix}

\end{document}